\pdfoutput=1
\documentclass[11pt]{article}

\usepackage[]{acl}
\usepackage{graphicx}
\usepackage{times}
\usepackage{latexsym}
\usepackage{tabularx}
\usepackage{booktabs}
\usepackage{multirow}
\usepackage{caption}
\usepackage{subcaption}
\usepackage{multicol}
\usepackage{amsmath}
\usepackage{amsfonts}
\usepackage{bm}
\usepackage{multirow}
\usepackage{colortbl}
\usepackage[T1]{fontenc}
\usepackage{graphicx}
\usepackage{array}
\usepackage{longtable}
\usepackage{bbding}

\usepackage[utf8]{inputenc}

\usepackage{microtype}

\title{Unsupervised Multimodal Clustering for \\Semantics Discovery in Multimodal Utterances}

\author{
  Hanlei Zhang\textsuperscript{\rm 1}, Hua Xu\textsuperscript{\rm 1}\thanks{\quad Hua Xu is the corresponding author.}, Fei Long\textsuperscript{\rm 1}, Xin Wang\textsuperscript{\rm 1,2,3}, Kai Gao\textsuperscript{\rm 2}\\
  \textsuperscript{\rm 1}State Key Laboratory of Intelligent Technology and Systems, \\ 
Department of Computer Science and Technology, Tsinghua University,\\
  \textsuperscript{\rm 2}School of Information Science and Engineering, Hebei University of Science and Technology\\
  \textsuperscript{\rm 3}Samton (Jiangxi) Technology Development Co.,Ltd, Nanchang 330036, China\\
    \texttt{zhang-hl20@mails.tsinghua.edu.cn, xuhua@tsinghua.edu.cn}\\
    }

\begin{document}
\maketitle
\begin{abstract}
Discovering the semantics of multimodal utterances is essential for understanding human language and enhancing human-machine interactions. Existing methods manifest limitations in leveraging nonverbal information for discerning complex semantics in unsupervised scenarios. This paper introduces a novel unsupervised multimodal clustering method (UMC), making a pioneering contribution to this field. UMC introduces a unique approach to constructing augmentation views for multimodal data, which are then used to perform pre-training to establish well-initialized representations for subsequent clustering. An innovative strategy is proposed to dynamically select high-quality samples as guidance for representation learning, gauged by the density of each sample's nearest neighbors. Besides, it is equipped to automatically determine the optimal value for the top-$K$ parameter in each cluster to refine sample selection. Finally, both high- and low-quality samples are used to learn representations conducive to effective clustering. We build baselines on benchmark multimodal intent and dialogue act datasets. UMC shows remarkable improvements of 2-7\% scores in clustering metrics over state-of-the-art methods,  marking the first successful endeavor in this domain. The complete code and data are available at \url{https://github.com/thuiar/UMC}.
\end{abstract}

\section{Introduction}
Discovering the semantics of dialogue utterances in unsupervised multimodal data requires integrating various modalities (i.e., text, video, and audio) to effectively mine the complicated semantics inherent in multimodal language. Conventional methods for semantics discovery typically focus solely on the text modality with clustering algorithms~\cite{zhang2021supporting,USNID}, failing to leverage the rich multimodal information in the real world (e.g., body language, facial expressions, and tones). 

\begin{figure}[t!]\small
	\centering  
	\includegraphics[scale=.45]{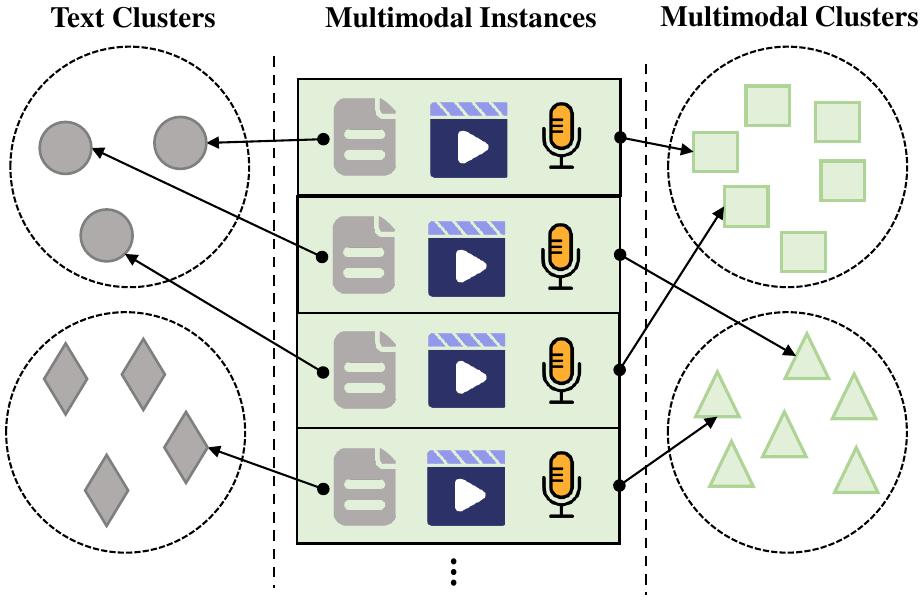}
	\caption{\label{example} Text-only clustering deviates from real multimodal utterance semantics, highlighting the need of multimodal information in semantics discovery.}
 \vspace{-0.1cm}
\end{figure}

However, we argue that non-verbal modalities (i.e., video and audio) also play a critical role when performing unsupervised clustering. Taking Figure~\ref{example} as an example, relying solely on textual information yields clustering results that differ from the ground truth of multimodal cluster allocations (a detailed analysis on real-world examples is available in Appendix~\ref{appendix:example}), suggesting that non-verbal modalities can
provide useful cues for semantics discovery. Moreover, effectively capturing multimodal interactions can yield more powerful and robust representations, thereby better addressing the challenges of  ambiguous intent-cluster boundaries found in text-based clustering (see Section~\ref{section: visualization} and Appendix~\ref{appendix:representation_visualization}). Discovering multimodal utterance semantics holds significant promise for a variety of applications, including video content recommendation, efficient multimodal data annotation, and virtual human technologies (detailed in Appendix~\ref{appendix:applications_future}). 

Understanding semantics in multimodal utterances has attracted much attention with the boom in multimodal language analysis~\cite{MELD,saha-etal-2021-towards_,zhang2022mintrec}. For example,~\citet{saha-etal-2021-towards_} annotated multimodal dialogue act (DA) labels on two popular multimodal multi-party conversational datasets~\cite{busso2008iemocap,MELD} and performed DA recognition using attention sub-networks build upon modality encoders.~\citet{zhang2022mintrec} pioneered multimodal intent analysis, introducing a new dataset with multimodal intent labels and establishing baselines with three multimodal fusion methods~\cite{tsai2019multimodal,rahman2020integrating,hazarika2020misa}. However, these works remain restricted within supervised tasks, i.e., the training target for each piece of data is known, which is not applicable in unsupervised scenarios.

In contrast, semantics discovery is an emerging field in NLP. It fundamentally operates as a clustering task and has seen the development of many unsupervised~\cite{cheung2012sequence,padmasundari2018intent,haponchyk2018supervised,USNID} and semi-supervised~\cite{lin2020discovering,Zhang_Xu_Lin_Lyu_2021,zhang2022new,zhou-2023-proba} methods. However, these methods are primarily designed for the text-only modality and lack proficiency in handling the diverse modalities encountered in real-world scenarios. Thus, there is a lack of multimodal clustering methods for discovering utterance semantics, posing two challenges: (1) determining how to leverage information from nonverbal modalities to complement the text modality in clustering and (2) devising ways to fully exploit multimodal unlabeled data to learn clustering-friendly representations.

\begin{figure*}[!ht]
    \centering
    \includegraphics[scale=0.95]{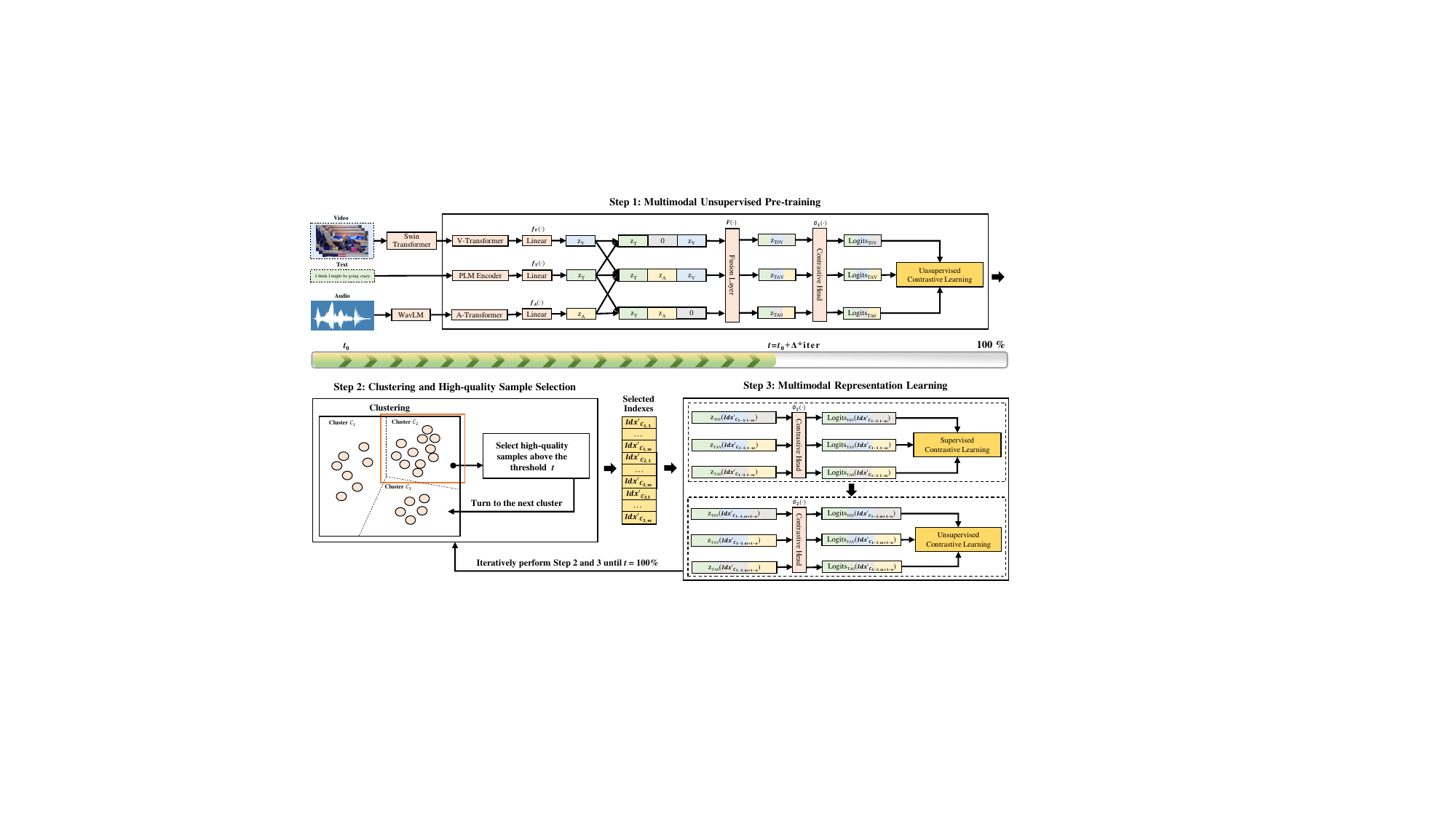}
    \caption{Overview of our proposed unsupervised multimodal clustering algorithm UMC.}
    \label{framework}
    % \vspace{-0.cm}
\end{figure*}

To address these challenges, we introduce UMC, a novel unsupervised multimodal clustering algorithm for semantics discovery, as shown in Figure~\ref{framework}. We utilize the capabilities of the pre-trained language model~\cite{BERT} to process text data. For the video and audio modalities, deep features are initially extracted using powerful backbones from computer vision and speech signal processing. Two transformer encoders are then employed to capture the deep semantics of these features. The text modality is designated as the anchor, guiding the learning of the other modalities. For this purpose, we concatenate features from all three modalities and mask the video or audio features with zero vectors, creating two sets of positive augmentation views. These multimodal representations and their augmentations are applied to an unsupervised contrastive loss, yielding well-initialized representations for subsequent process. 

To fully mine the semantic similarities among unsupervised multimodal data, we introduce a novel strategy that initially selects high-quality samples. This strategy employs a dynamic sample selection threshold $t$, aiming to select the highest-quality $t$ percent of samples in each iteration for training. This selection is based on a unique mechanism that calculates the density of each sample within its respective cluster and ranks them accordingly. Besides, an evaluation process is designed to automatically determine the optimal parameters for the top-$K$ nearest neighbors from a set of candidates. After selecting high-quality samples, we propose a sequential process for multimodal representation learning. This process begins by learning from high-quality samples using supervised contrastive loss and then refines the remaining low-quality samples using unsupervised contrastive loss. This two-step approach promotes beneficial intra-class and inter-class relations among high-quality samples while pushing apart low-quality samples, thereby generating representations conducive to clustering. The entire process is repeated until the sample selection threshold $t$ is met.

We summarize our contributions as follows: 

In this work, we make a pioneering contribution by formulating the challenging multimodal semantics discovery task. To solve this problem, we first introduce a novel method for constructing positive augmentations for multimodal data, effectively leveraging non-verbal modalities for unsupervised pre-training, which provides a good initialization for unsupervised clustering. 

Then, we propose a new clustering algorithm, UMC, which features an innovative high-quality sample selection strategy and a sequential representation learning method between high- and low-quality samples, resulting in excellent performance across both single and multimodal modalities. 

Finally, we establish baselines using benchmark multimodal intent and dialogue datasets. Extensive experiments show that the proposed UMC outperforms state-of-the-art clustering algorithms by a notable margin of 2-7\% scores in standard clustering metrics. To the best of our knowledge, this is the first successful attempt at leveraging multiple modalities for unsupervised clustering, marking a substantial advancement in this area. 

\section{Related Works}

\subsection{Unsupervised Clustering}
Unsupervised clustering is fundamental in machine learning. Classic clustering methods like K-Means~\cite{macqueen1967some} and Agglomerative Clustering~\cite{gowda1978agglomerative} iteratively assign clusters until convergence based on features. Deep clustering methods, like DEC~\cite{xie2016unsupervised} and DCN~\cite{yang2017towards}, enhance this process by jointly clustering and feature learning, employing stacked autoencoders~\cite{vincent2010stacked}. DeepCluster~\cite{caron2018deep} uses cluster assignments as guidance for feature learning. 

Recent methods using contrastive learning~\cite{chen2020simple} have achieved the state-of-the-art performance. For instance, SCCL~\cite{zhang2021supporting} combines instance-level contrastive learning with cluster refinement from target distributions. CC~\cite{kumar-etal-2022-intent_} optimizes contrastive losses at both instance and cluster levels to generate clustering-friendly representations. However, these methods focus on merely the single text or image modality and fall short with multimodal data. MCN~\cite{chen2021multimodal} is tailored for multimodal clustering, learning a unified representation from all modalities and applying cross-modal contrastive losses during clustering. However, MCN struggles with complex utterance semantics.

\subsection{Intent  Discovery}
Intent discovery is a key challenge in NLP, with numerous clustering methods developed to address it. Early methods~\cite{hakkani2015clustering,haponchyk2018supervised} use weakly supervised signals to aid in clustering but struggle to capture  the high-level semantics in text. Recent methods~\cite{lin2020discovering,Zhang_Xu_Lin_Lyu_2021,mou-2022-disentangled,zhang2022new,mou-2023-decoupling,zhou-2023-proba,shi-2023-diffusion} exploit limited labeled data to guide the feature learning process for clustering. 

However, these methods suffer a substantial decrease in performance in totally unsupervised scenarios. USNID~\cite{USNID} proposes a novel centroid-guided mechanism with a pre-training strategy, achieving significant improvements over previous methods. Yet, USNID also falls short in handling multimodal data. See Appendix~\ref{appendix:additional_related_works} for more related works on multi-view clustering and multimodal language analysis.

\section{Problem Formulation}
For the task of multimodal semantics discovery, we are provided with a multimodal intent or dialogue act dataset $\mathcal{D}_{\textrm{mm}}=\{(\boldsymbol{s}_{i}^{\text{T}}, \boldsymbol{s}_{i}^{\text{A}}, \boldsymbol{s}_{i}^{\text{V}})|y_{i} \in \mathcal{I}, i=1,..., N\}$, where each $i^{\textrm{th}}$ instance $\boldsymbol{s}_{i}$ contains multimodal utterances, including $\boldsymbol{s}_{i}^{\text{T}}$, audio $\boldsymbol{s}_{i}^{\text{A}}$, and video $\boldsymbol{s}_{i}^{\text{V}}$. Here, $N$ represents the total number of instances. The ground-truth label $y_{i}$, belonging to the set of intent or dialogue act classes $\mathcal{Y}=\{y_i\}^{K_{\mathcal{Y}}}_{i=1}$, remains unseen during training and validation and is only available during testing. The number of classes is denoted by $K_{\mathcal{Y}}$. 

The objective is to learn a multimodal neural network $\mathcal{F}(\cdot)$ capable of obtaining multimodal representations $\boldsymbol{z}$ conducive to clustering. These representations are subsequently employed to divide the set $\{\boldsymbol{s}_{i}\}^{N}_{i=1}$ into $K_{\mathcal{Y}}$ groups.

\section{Methodologies}
\subsection{Multimodal Representation}

To obtain multimodal representations, we first extract deep features from text, video, and audio modalities. For text, we employ the pre-trained language model (PLM), BERT~\cite{BERT} as the encoder, fine-tuning it on the text inputs $\boldsymbol{s}^{\text{T}}$. The initial [CLS] token embedding, $\boldsymbol{x}_{\text{T}} \in \mathbb R^{D_{\text{T}}}$, serves as the sentence-level representation, where $D_{\text{T}}$ is the feature dimension of 768. We then incorporate a linear layer, represented as $f_{\text{T}}(\cdot)$, yielding $\boldsymbol{z}_{\text{T}} \in \mathbb R^{D_{H}}$. Here, $H$ indicates a dimensionally reduced space, enhancing computational efficiency and accentuating primary features.

For non-verbal modalities, we use semantically rich features as inputs as suggested in~\cite{EMOTyDA,zhang2022mintrec}. For video, we 
employ the Swin Transformer~\cite{liu2021swin} to extract video feature representations $\boldsymbol{x}_{\text{V}} \in \mathbb R^{L_{\text{V}} \times D_{\text{V}}}$ at the frame level from the video inputs $\boldsymbol{s}^{\text{V}}$. Here, $L_{\text{V}}$ represents the video length, and $D_{\text{V}}$ is the feature dimension of 1024. For audio $\boldsymbol{s}^{\text{A}}$, we first extract audio waveforms as in~\cite{zhang2022mintrec} and then use the WavLM~\cite{chen2022wavlm} to obtain features $\boldsymbol{x}_{\text{A}} \in \mathbb R^{L_{\text{A}} \times D_{\text{A}}}$. Here, $L_{\text{A}}$ and $D_{\text{A}}$ denote the audio length and feature dimension of 768, respectively. Unsupervised multimodal clustering can benefit from these two powerful non-verbal features extracted from the Swin Transformer and WavLM models. A comparison between them and other multimodal features is shown in Appendix~\ref{appendix:effect_of_multimodal}. 

For both audio and video modalities, initially introduce a linear layer $f_M(\cdot)$ in alignment with the text modality. Subsequently, we apply the multi-headed attention mechanism with the Transformer~\cite{Transformer} encoder, adeptly capturing intricate semantic relationships and temporal nuances. Eventually, in line with~\cite{tsai2019multimodal}, the last sequence elements are employed to derive the sentence-level representation $\boldsymbol{z}_{M}$:
\begin{align}
    \boldsymbol{z}_{M}=\text{Transformer}(f_{M}(\boldsymbol{x}_{M}))[-1],
\end{align}
where $M \in \{\text{A}, \text{V}\}$, and $\boldsymbol{z}_{M} \in \mathbb R^{D_{H}}$.

Following this, we concatenate the representations  $\boldsymbol{z}_{\text{T}}$, $\boldsymbol{z}_{\text{A}}$, and $\boldsymbol{z}_{\text{V}}$ and pass them through a non-linear fusion layer, denoted as $\mathcal{F}: \mathbb{R}^{3D_{H}} \rightarrow \mathbb{R}^{D_{H}}$. This layer is designed to learn cross-modal interactions, yielding the combined representation $\boldsymbol{z}_{\text{TAV}} \in \mathbb{R}^{D_H}$:
\begin{align}
\boldsymbol{z}_{\text{TAV}} = \mathcal{F}(\text{Concat}( \boldsymbol{z}_{\text{T}}, \boldsymbol{z}_{\text{A}}, \boldsymbol{z}_{\text{V}})),
\label{eq2}
\end{align}
where $\mathcal{F}$ is defined as $W_{1} \sigma_{\text{GELU}}(\text{Dropout}(\cdot)) + b_{1}$. Here, $\sigma_{\text{GELU}}$ represents the GELU activation function, and $W_{1}$ and $b_{1}$ are the corresponding weight and bias matrices, respectively. Subsequently, we employ $\boldsymbol{z}_{\text{TAV}}$ and its augmentations for further clustering and representation learning.

\begin{figure*}[!ht]
    \centering
    \includegraphics[scale=1.2]{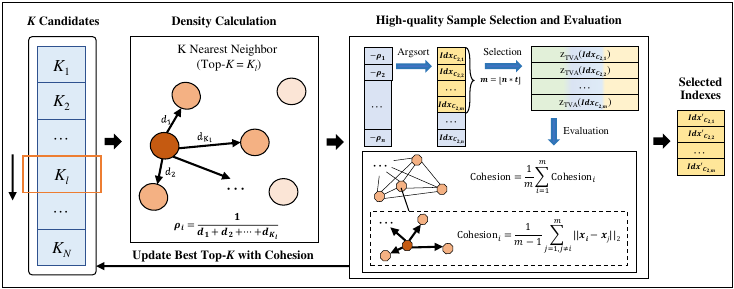}
    \caption{Pipeline of the high-quality sample selection mechanism.}
    \label{module}
\end{figure*}

\subsection{Multimodal Unsupervised Pre-training}
\label{mm_unsup}
Effective pre-training strategies can provide well-initialized representations conducive to clustering~\cite{USNID}. Unsupervised contrastive learning~\cite{chen2020simple} has emerged as an effective approach for unsupervised clustering~\cite{zhang2021supporting,li2021contrastive}.  It pushes apart samples and makes them distribute uniformly in the feature space while capturing implicit similarity relations between augmentations. However, existing methods often fall short in providing effective augmentations for multimodal data. In this work, we introduce a novel method of non-verbal modality masking to address this gap.

Given the predominant role of the text modality in intent analysis, we retain it as the core modality and mask either the video or audio modality for data augmentation. For the $i^{\text{th}}$ sample $\boldsymbol{z}_{\text{TAV}, i}$ in a minibatch of $B$ samples, either the video or audio modality is replaced with zero vectors. Eq.~\ref{eq2} is used to derive $\boldsymbol{z}_{\text{TA0}, i}$ and $\boldsymbol{z}_{\text{T0V}, i}$ as positively augmented samples. For each positive pair ($i$, $j$)  among the generated $3B$ augmented samples, we apply the multimodal unsupervised contrastive learning loss:
\begin{align}
    &\mathcal{L}^{\textrm{mucl}}_{i, j} =\nonumber \\ 
    &-\log \Bigg( \frac{\exp(\textrm{sim}(\phi_1(\boldsymbol{z}_i), \phi_1(\boldsymbol{z}_{j}))/\tau_1)}{\sum_{k} \mathbb{I}_{[k \neq i]} \exp(\textrm{sim}(\phi_1(\boldsymbol{z}_i), \phi_1(\boldsymbol{z}_{k}))/\tau_{1})} \Bigg),
\label{ucl_formula}
\end{align}
where $\boldsymbol{z}_i \in \{\boldsymbol{z}_{\text{TAV}, i}, \boldsymbol{z}_{\text{TA0}, i}, \boldsymbol{z}_{\text{T0V}, i}\}$, $\textrm{sim}(\cdot)$ refers to the dot product operation on two L2-normalized vectors, and $\phi_1(\cdot)$ is a non-linear layer with ReLU activation, serving as the contrastive head. The parameter $\tau_{1}$ represents the temperature, and $\mathbb{I}_{[\cdot]}$ is the indicator function, outputting 1 if and only if $j=i$, and 0 otherwise. 

By masking the video or audio modality with zero vectors, the model can focus on learning the implicit similarities in the shared modalities among positive pairs (i.e., text and video, text and audio, and text alone). This further encourages the model to capture intricate relationships and leverage complementary information across modalities.

% $\boldsymbol{z}_i$ and $\boldsymbol{z}_{i}^{\prime}$ are derived from the positive pairs in $\{\boldsymbol{z}_{\text{TAV},i}, \boldsymbol{z}_{\text{T0V}, i}, \boldsymbol{z}_{\text{TA0}, i}\}$. 
\subsection{Clustering and High-Quality Sample Selection}

After pre-training, we employ the representations $\boldsymbol{z}_{\text{TAV}}$ to perform clustering. Specifically, we adopt the K-Means++ algorithm~\cite{arthur2007k} for this task due to its advanced initial centroid selection technique that improves convergence over standard K-Means. 

However, we observe that the cluster assignments obtained directly from K-Means++ are insufficiently high-quality to guide the learning of multimodal representations. To address this, we introduce a strategy to incrementally incorporate high-quality samples into the learning process. This is achieved through a curriculum-based method, where we progressively adjust the sample selection threshold, $t$, dictating the proportion of selected samples from each cluster for a given training iteration. The threshold $t$ is linearly updated as follows:
\begin{equation}
    t = t_0 + \Delta \cdot \text{iter},
    \label{eqt}
\end{equation}
where $t,t_0 \in [0,1]$, $t_0$ is the initial threshold set to 0.1 (see Appendix~\ref{appendix:t_0_selection} for a detailed discussion), $\text{iter}$ is the iteration index within the epoch, and $\Delta$ is a preset positive increment, applied after each epoch. 

To further refine clustering performance, we incorporate the centroid inheritance strategy as proposed in~\cite{USNID}. Specifically, K-Means++ is utilized only during the first training iteration. In subsequent iterations, the cluster centroids from the previous iteration are inherited as initial centroids. This approach effectively leverages historical clustering information to guide and improve current clustering results. 

Then, we need to identify high-quality samples for representation learning. We introduce a novel mechanism for selecting high-quality samples, as depicted in Figure~\ref{module}. This mechanism comprises two main steps: density calculation and high-quality sample selection and evaluation. 

\subsubsection{Density Calculation}
To discern high-quality samples within each cluster, we propose using density as the criterion. The underlying intuition is that high-quality samples are likely to exhibit high local density, whereas low-quality, anomalous, or falsely clustered data are expected to have low local density. For the $i^{\textrm{th}}$ sample, we compute its density, $\rho_{i}$, as the reciprocal of the average distance between $\boldsymbol{z}_{\text{TAV}, i}$ and its top-$K$ nearest neighbors:
\begin{align}
\label{eq_rou}
\rho_i = \frac{K_{\text{near}}}{\sum_{j=1}^{K_{\text{near}}} d_{ij}},
\end{align}
where $K_{\text{near}}$ denotes the number of top-$K$ nearest neighbors. $d_{ij}$ represents the Euclidean distance between the $i^{\textrm{th}}$ sample and its $j^{\textrm{th}}$ nearest neighbor. 

\subsubsection{High-Quality Sample Selection and Evaluation}
\label{sample_selection}
After calculating the density of each sample in each cluster, we rank them based on their densities in descending order. Specifically, for each sample in the $k^{\textrm{th}}$ cluster $C_k$ with a density of $\rho_i$, we compute a sorted index list $Idx_{C_k}$ as follows:
\begin{align}
\label{idx}
Idx_{C_k} = \operatorname{argsort}(-[\rho_{1}, \rho_{2}, \cdots, \rho_{n}]),
\end{align}
where $\operatorname{argsort}$ yields the indices that sort the densities in ascending order of the negative values, and $n$ represents the number of samples in $C_k$. The high-quality samples are selected based on the highest densities. The number of selected highest-density samples has a proportion in cluster $C_k$ above the threshold $t$. Let $m=\lfloor n * t \rfloor$, the chosen samples are denoted as: $\boldsymbol{z}_{\text{TAV}}(Idx_{C_{k,1}}), \ldots, \boldsymbol{z}_{\text{TAV}}(Idx_{C_{k,m}})$. $\boldsymbol{z}_{\text{TAV}}(Idx_{C_{k,i}})$ is the $i^{\textrm{th}}$ selected sample feature in cluster $C_k$, based on the ordered density indices.

Considering that real-world data might not exhibit a uniform distribution across each class, assigning a fixed $K_{\text{near}}$ to every cluster could compromise the precision of density calculations, subsequently affecting the selection of high-quality samples. To address this, we introduce an innovative method to automatically select the optimal $K_{\text{near}}^{k}$ for each cluster $C_k$. Initially, we provide a candidate set $\{K_{\text{near}, q}^{k}\}_{q=1}^{u}$, uniformly sampled based on the cluster size $|C_k|$. Specifically, $K_{\text{near}, q}^{k}$ is defined as:
\begin{align}
    K_{\text{near}, q}^{k}=\lfloor|C_k|\cdot(L+\Delta^{\prime}\cdot(q-1)) \rfloor,
    % u = \frac{(U-L)\cdot|C_k|}{\Delta^{\prime}}+1,
\end{align}
where $L$ is the lower proportion bound with the constraint of $0\leq L\leq 1$, $\Delta^{\prime}$ is a fixed interval, and $u$ is the number of candidates. Then, for each candidate $K_{\text{near}, q}^{k}$, we use Eq.~\ref{idx} to compute sorted indices $Idx_{C_{k}}^q$ and select a subset $C_{k}^{q}$ with top-$m$ samples. The quality of $C_{k}^{q}$ is gauged through the cluster cohesion metric,  measuring intra-cluster similarity. The cohesion of $C_{k}^{q}$ is defined as:
\begin{align}
    \text{coh}(C_k^{q})&=\sum_{i=1}^{m}\text{coh}(C_{k,i}^{q}),\\
    \text{coh}(C_{k,i}^{q})&=\\ \nonumber
   \frac{1}{m-1}&\sum_{j=1,j\neq i}^{m}d(\boldsymbol{z}_{\text{TAV}}(Idx_{C_{k,i}^q}), \boldsymbol{z}_{\text{TAV}}(Idx_{C_{k,j}^q})),
\end{align}
where $m$ is the previously defined number of chosen samples, $d(\cdot)$ represents the Euclidean distance. The cohesion score can effectively capture the feature compactness and reflect the cluster quality. The optimal selected candidate index $q_{\text{opt}}$ is calculated as:
\begin{align}
    q_{\text{opt}}=\underset{q}{\operatorname{argmin}}\left\{\text{coh}(C_k^{q})\right\}.
\end{align} 

That is, the optimal $K_{\text{near},q_{\text{opt}}}^k$ is selected by the candidate with the highest cluster cohesion score. Subsequently, we use $\{K_{\text{near},q_{\text{opt}}}^k\}_{k=1}^{K_{\mathcal{Y}}}$ to obtain the selected high-quality indices ${Idx}^{\prime}=\{Idx_{C_{k}}^{q_{\text{opt}}}\}_{k=1}^{K_{\mathcal{Y}}}$ with  Eq.~\ref{eq_rou} and~\ref{idx}. These indices are then employed to select high-quality samples for subsequent representation learning.

\subsection{Multimodal Representation Learning}
The high-quality samples identified by the selected indices $Idx'$ tend to have more reliable pseudo-labels, so we employ them as a guiding set to facilitate the learning of friendly representations for clustering. We aim to leverage these samples to capture high-level similarity relations between pairwise samples. To achieve this, we introduce the multimodal supervised contrastive loss: 
\begin{align}
     \label{scl}
        &\mathcal{L}_{i}^{\textrm{mscl}}=\nonumber \\ 
        &\frac{-1}{|{P}(i)|}\sum_{p\in\mathcal{P}(i)}\log\frac{\exp(\textrm{sim}(\boldsymbol{l}_{i},\boldsymbol{l}_{p})/\tau_2)}{\sum_{j}\mathbb{I}_{[j\neq i]}\exp(\textrm{sim}(\boldsymbol{l}_{i},\boldsymbol{l}_{j})/\tau_2)},
\end{align}
where $\boldsymbol{l}_{i}=\phi_2(\boldsymbol{z}_{i})$, and $\phi_2$ is a non-linear layer with ReLU activation, consistent with Eq.~\ref{ucl_formula}. Here, we perform the same data augmentation techniques as in Section~\ref{mm_unsup},  and $\boldsymbol{l}_i \in \{\boldsymbol{l}_{\text{TAV},i}, \boldsymbol{l}_{\text{T0V}, i}, \boldsymbol{l}_{\text{TA0}, i}\}$. $\tau_2$ denotes the temperature parameter. $P(i)$ is the set of indices for the augmented samples that share the same classes with $\boldsymbol{l}_{i}$. With this loss,  each sample can learn not only from its respective augmentations but also learn from the  clustering information derived from high-quality pseudo-labels.

Conversely, low-quality samples are prone to erroneous clustering, where
dissimilar samples may be grouped into the same class. This misgrouping can disrupt the integrity of the clustering process. To mitigate this issue, we propose the application of an unsupervised contrastive loss to these samples. This loss function is designed to increase the separation between distinct low-quality samples, thereby encouraging a more uniform distribution in the feature space, as supported by~\cite{zhang2021supporting}. Specifically, we use Eq.~\ref{ucl_formula}, replacing $\phi_1$ with $\phi_3$, and apply this modified equation to the remaining samples in the training set, excluding those with selected indices $Idx'$. 

In our approach, we sequentially apply multimodal supervised contrastive learning to high-quality samples and unsupervised contrastive learning to low-quality samples. This two-step strategy is crafted to concurrently enhance multimodal representation learning and clustering process. The training phase concludes when the sample selection threshold $t$ (as defined in Eq.~\ref{eqt}) reaches 100\%. During the inference stage, we utilize the well-trained model to extract $\boldsymbol{z}_{\text{TAV}}$ and subsequently employ the K-Means++ algorithm for prediction.

\begin{table}[t!]\small
    \centering
    \begin{tabular}{@{} lcccccc @{}}
        \toprule
        Datasets & \#C & \#U & \#Train & \#Test \\
        \midrule
        MIntRec & 20 & 2,224 & 1,779 & 445 \\ 
        MELD-DA & 12 & 9,988 & 7,990 & 1,998   \\
        IEMOCAP-DA & 12 & 9,416 & 7,532 & 1,884 \\ 
        \bottomrule
    \end{tabular}
     \caption{ \label{datasets}  Statistics of MIntRec, MELD-DA, IEMOCAP-DA datasets. \# indicates the total number of sentences. \#C and \#U denote the number of classes and utterances.}
\end{table}

\section{Experiments}
\subsection{Datasets}
We use MIntRec, MELD-DA, and IEMOCAP-DA as benchmark datasets for the multimodal semantics discovery task. The rationale for using these datasets is that the defined intents or dialogue acts typically exhibit a variety of distinct sentence-level semantics and possess properties of uncertainty in the open world, making them suitable for discovery in unsupervised scenarios. Detailed statistics of the three datasets are presented in Table~\ref{datasets}, with further information on dataset specifics and their splits available in Appendix~\ref{appendix:dataset_specifications}.

\subsection{Baselines}

We compare UMC with the state-of-the-art unsupervised clustering methods from both NLP and CV, as well as multimodal clustering methods. The TEXTOIR platform~\cite{zhang-etal-2021-textoir_} is used to reproduce the methods in NLP. Detailed descriptions of the baselines are follows: 

\textbf{SCCL}~\cite{zhang2021supporting}: It jointly optimizes clustering and instance-level contrastive learning losses. The learning rate is set to 3e-5.

\textbf{CC}~\cite{li2021contrastive}: It employs dual non-linear heads to independently optimize instance-level and cluster-level contrastive learning losses.  The learning rate is set to 3e-5.

\textbf{USNID}~\cite{USNID}: It  performs strong data augmentation by randomly erasing words in a sentence. It also introduces a centroid-guided clustering mechanism to construct high-quality supervised signals for representation learning . 

\textbf{UMC (Text)}: This UMC variant excludes video and audio modalities during clustering. Unlike UMC, which uses multimodal augmentations, here we apply \textit{dropout} twice to generate positive augmentations for contrastive learning. 

\textbf{MCN}~\cite{chen2021multimodal}: It employs an online K-Means algorithm to dynamically determine cluster centers and periodically update them. However, we find its performance drops with online clustering. Thus, we modify it to perform K-Means on the full dataset to ensure optimal performance.

\subsection{Evaluation Metrics}
Following~\cite{6832486,saxena2017review}, we use four standard clustering metrics to evaluate the clustering performance, including Normalized Mutual Information (NMI),  Accuracy (ACC), Adjusted Rand Index (ARI), and Fowlkes-Mallows Index (FMI). Details can be found in Appendix~\ref{em}. 

\begin{table}[!t] \small
    \centering
     \scalebox{0.9}{
    \begin{tabular}{@{\extracolsep{0.0001pt}}ll|lccccc}
    \toprule
       & Methods & NMI & ARI & ACC & FMI &  Avg. \\
        \midrule 
         \multirow{6}{*}[1ex]{\rotatebox[origin=c]{90}{MIntRec}}
        & SCCL & 45.33 & 14.60 & 36.86 &  24.89 &  30.42  \\
       & CC & 47.45 & 22.04 & 41.57 & 26.91 & 34.49  \\
        & USNID & 47.91 & 21.52 & 40.32 & 26.58 &  34.08 \\    
        & MCN & 18.24 & 1.70 & 16.76 & 10.32 &  11.76 \\
        \cmidrule{2-7}
        & UMC (Text) & 47.99 & 21.33 & 42.00 & 26.33 & 34.41 \\
        & UMC & \textbf{49.46} & \textbf{24.79} & \textbf{44.00} & \textbf{29.48} & \textbf{36.93} \\
        \midrule
        \midrule
         \multirow{6}{*}[1ex]{\rotatebox[origin=c]{90}{M-DA}}
        & SCCL & 22.42 & 14.48 & 32.09 & 27.51 &  24.13  \\
        & CC & 23.03 & 13.53 & 25.13 & 24.86 &  21.64 \\
        & USNID & 20.80 & 12.16 & 24.07 & 23.28 & 20.08  \\
        & MCN & 8.34 & 1.57 & 18.10 & 15.31 &  10.83 \\
        \cmidrule{2-7}
        & UMC (Text) & 20.33 & 17.90 & 32.54 & 31.29 & 25.52 \\ 
        & UMC & \textbf{23.18} & \textbf{21.34} & \textbf{35.58} & \textbf{34.39} & \textbf{28.62} \\
        \midrule
        \midrule
        \multirow{6}{*}[1ex]{\rotatebox[origin=c]{90}{I-DA}}
        & SCCL& 21.90 & 10.90 & 26.80 & 24.14 &  20.94  \\
        & CC & 23.59 & 12.99 & 25.86 & 24.42 &  21.72  \\
        & USNID & 22.19 & 11.92 & 27.35 & 23.86 &  21.33  \\
        & MCN & 8.12 & 1.81 & 16.16 & 14.34  & 10.11 \\
        \cmidrule{2-7}
        & UMC (Text) & 20.99 & 19.05 & 32.40 & 31.39 & 25.96 \\  
        & UMC & \textbf{24.60} & \textbf{21.43} & \textbf{36.36} & \textbf{34.42} & \textbf{29.21} \\
        \bottomrule
    \end{tabular}}
     \caption{\label{results-m}Results on  MIntRec, MELD-DA (M-DA), and IEMOCAP-DA (I-DA) datasets. }
\end{table}

\subsection{Experimental Setup}
For the text modality, we utilize the pre-trained BERT model from the Huggingface Transformers library~\cite{wolf2020transformers} and optimize it using the AdamW~\cite{loshchilov2017decoupled} optimizer. It is important to note that all the baselines utilize the same backbone for each of the three modalities for a fair comparison. In our experiments, the multimodal data employed for pre-training and training adhere to a consistent distribution and characteristics, and no external data is used for pre-training. 

We configure the sequence lengths $L_{\text{T}}, L_{\text{V}}, L_{\text{A}}$ for MIntRec, MELD-DA, and IEMOCAP-DA datasets to (30, 230, 480), (70, 250, 520), and (44, 230, 380), respectively. The threshold $t$ is incremented by $\Delta$ of 0.05. For the selection of optimal $K_{\text{near}}$, we configure $L=0.1$, $\Delta^{\prime}=0.02$, and $u=10$. The learning rates are 2e-5 and (3e-4, 2e-4, 5e-4) for pre-training and training stages of MIntRec, MELD-DA, and IEMOCAP-DA datasets. The temperature parameters $\tau_1$, $\tau_2$, and $\tau_3$ are set at (0.1,7.5,0.9), (0.1, 4, 4) and (0.1, 4, 8) for these datasets, respectively. A detailed hyper-parameter sensitivity analysis is provided in Appendix~\ref{appendix:sensitivity_analysis}. We use a training batch size of 128 and report an average performance over five random seeds of 0-4.

\begin{table}[!t] \small
    \centering
     \scalebox{0.8}{
    \begin{tabular}{@{\extracolsep{0.01pt}}c|l|cccccc}
    \toprule
         & Methods & NMI & ARI & ACC & FMI  & Avg.  \\
        \midrule
        \multirow{8}{*}[1ex]{\rotatebox[origin=c]{90}{MIntRec}}
        & w/o Step 1 & 39.15 & 14.24 & 31.91 & 19.51 & 26.20 \\
        & Random (Step 2) & 47.80 & 23.27 & 43.60 & 28.03 & 35.67 \\
        & SCL (Step 3) & 37.78 & 13.22 & 29.93 & 19.44 & 25.09 \\
        & Step 1$\&$K-Means++ & 37.62 & 12.32 & 32.00 & 17.66 & 24.90 \\
        & Step 1$\&$UCL & 47.08 & 22.02 & 43.95 & 26.84 & 34.97 \\
        & Step 1$\&$MSE & 24.57 & 3.86 & 18.61 & 10.38 & 14.35 \\
        & UMC & \textbf{49.46} & \textbf{24.79} & \textbf{44.00} & \textbf{29.48} & \textbf{36.93} \\
        \midrule
        \midrule
        \multirow{8}{*}[1ex]{\rotatebox[origin=c]{90}{M-DA}}
       & w/o Step 1 & 10.17 & 4.50 & 22.15 & 20.09 & 14.23 \\
        & Random (Step 2) & 20.49 & 17.35 & 31.73 & 30.45 & 25.01 \\
        & SCL (Step 3) & 14.08 & 8.55 & 26.43 & 22.48 & 17.88 \\
        & Step 1$\&$K-Means++ & 14.24 & 6.71 & 22.93 & 18.63 & 15.63 \\
        & Step 1$\&$UCL & 22.14 & 14.77 & 27.54 & 26.32 & 22.69 \\
        & Step 1$\&$MSE & 11.73 & 6.39 & 28.05 & 23.63 & 17.45 \\
        & UMC & \textbf{23.18} & \textbf{21.34} & \textbf{35.58} & \textbf{34.92} & \textbf{28.76} \\
        \midrule
        \midrule
        \multirow{8}{*}[1ex]{\rotatebox[origin=c]{90}{I-DA}}
        & w/o Step 1 & 9.72 & 3.86 & 24.25 & 19.87 & 14.43 \\
        & Random (Step 2) & 19.91 & 12.60 & 28.45 & 25.67 & 21.67 \\
        & SCL (Step 3) & 9.71 & 3.31 & 22.85 & 19.54 & 13.85 \\
        & Step 1$\&$K-Means++ & 11.17 & 4.38 & 20.01 & 16.58 & 13.04 \\
        & Step 1$\&$UCL & 22.35 & 14.48 & 29.13 & 26.58 & 23.14 \\
        & Step 1$\&$MSE & 10.60 & 3.11 & 21.16 & 18.32 & 13.30 \\
        & UMC & \textbf{24.60} & \textbf{21.43} & \textbf{36.36} & \textbf{34.42} & \textbf{29.21} \\
        \bottomrule
    \end{tabular}}
    \caption{\label{ablation} Ablation studies on the three datasets.}
\end{table}
\subsection{Results}

Table~\ref{results-m} shows the results on the multimodal semantics discovery task. 
Overall, UMC consistently outperforms all baseline methods across all datasets and evaluation metrics, achieving state-of-the-art performance. 
Specifically, compared to the best baseline in each metric, UMC achieves average performance improvements of 2.44\%, 4.49\%, and 7.49\% on MIntRec, MELD-DA, and IEMOCAP-DA, respectively. 
This demonstrates that our proposed high-quality sample selection and unsupervised learning framework can effectively assist the model in learning multimodal representations tailored for clustering. 
Notably, the only multimodal clustering baseline, MCN, exhibits poor performance, lagging behind UMC by a substantial margin of 17\% to 25\%. 
This disparity highlights the apparent limitations of computer vision-centric clustering methods when applied to multimodal language understanding tasks dominated by the discrete text modality.

UMC (Text), a unimodal variant of our method utilizing only the text modality, exhibits comparable or superior performance to existing SOTA methods across most clustering metrics. 
In particular, while UMC (Text) performs comparably to the strongest baseline on the MIntRec dataset, it yields significant FMI improvements of 3.78\% and 7.25\% on MELD-DA and IEMOCAP-DA, respectively. 
This underscores the critical role of high-quality sample selection and multi-stage optimization in both clustering and representation learning. 
Furthermore, compared to UMC (Text), UMC incorporates non-verbal modalities and demonstrates significant and consistent average improvements of 2.52\%, 3.10\%, and 3.25\% on the MIntRec, MELD-DA, and IEMOCAP-DA datasets, respectively. 
On the ARI metric specifically, UMC surpasses UMC (Text) by 3.46\%, 3.44\%, and 2.38\% on the three datasets, demonstrating that non-verbal modalities and our multimodal representation learning can more fully exploit non-linguistic cues to boost unsupervised data mining efficiency.

When examining individual metrics, UMC exhibits particularly pronounced improvements in ARI, ACC, and FMI compared to the baselines. 
Across the three datasets, UMC improves ARI by 2-8\%, ACC by 2-9\%, and FMI by 2-10\%. 
Compared to NMI, these metrics are more reflective of instance-level intra-cluster compactness and inter-cluster separation. 
These notable increases verify the capability of UMC to capture complex, fine-grained semantic differences. 
By dynamically introducing high-quality samples, our method adapts to complex environments in a curriculum-like (easy-to-hard) manner, leveraging high-quality pseudo-labels to guide representation learning. 
In contrast, although baselines like SCCL, CC, and USNID employ instance-level contrastive learning, they fail to achieve optimal performance due to the lack of high-quality sample guidance and non-verbal modality integration. 
A case study for error analysis is detailed in Appendix~\ref{appendix:error_analysis}.
\begin{figure}[!t]
    \centering
    \includegraphics[scale=0.145]{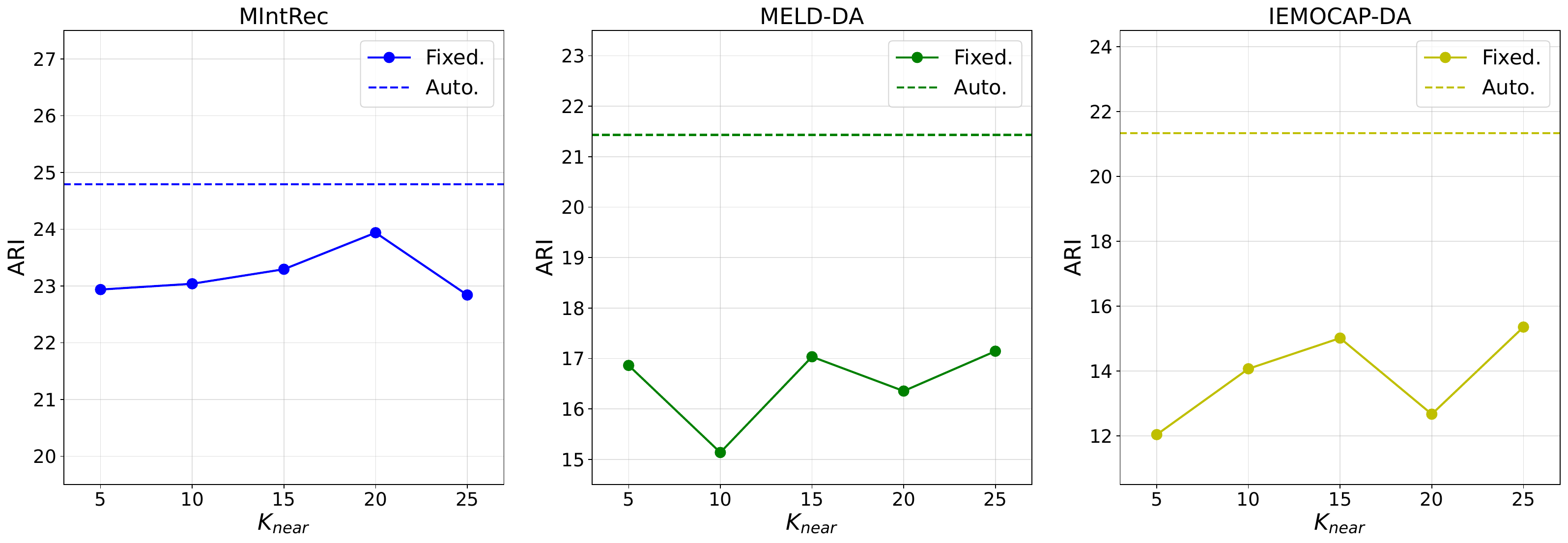}
    \caption{Automatic vs. fixed $K_{\text{near}}$ selection strategy.}
    \label{ablation_k}
    \vspace{-1.4cm}
\end{figure}

\begin{figure*}[!t]
    \centering
    \includegraphics[scale=0.15]{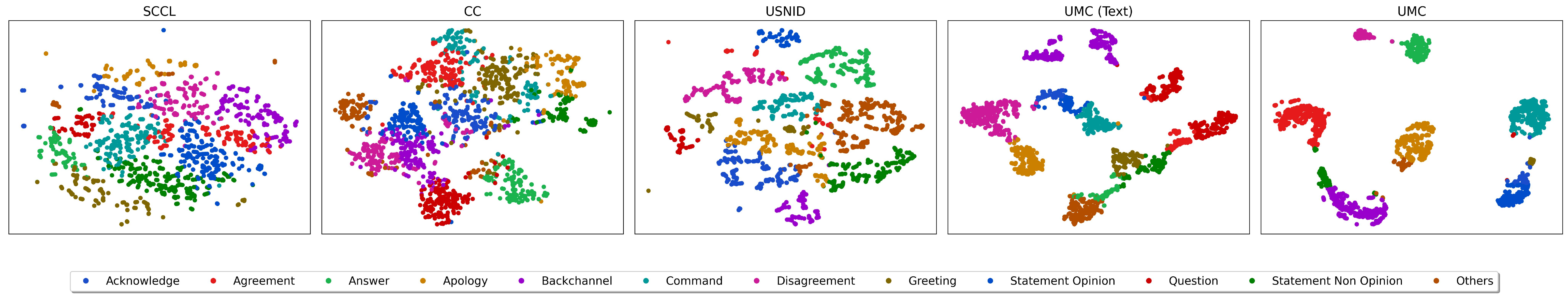}
    \caption{Visualization of representations on the IEMOCAP-DA dataset.}
    \label{tsne}
       % \vspace{-0.6cm}
\end{figure*}
\section{Discussion}
\subsection{Ablation Studies}

We conduct extensive ablation studies and show the results in Table~\ref{ablation}. (1) w/o Step 1: Removing Step 1 results in performance drops of 10-12\%, 13-16\%, and 12-17\% across the MIntRec, MELD-DA, and IEMOCAP-DA datasets, emphasizing the importance of our proposed non-verbal modality masking strategy in enhancing subsequent clustering. 

(2) Random (Step 2): To assess the impact of our high-quality sampling strategy in Step 2, we replace it with random sampling (i.e., randomly selecting the top-\textit{t} percent of samples from each cluster). This change leads to average score decreases of 1.26\%, 3.61\%, and 7.54\% on the clustering metrics, highlighting that carefully selected high-quality samples are pivotal in guiding the learning of multimodal representations. 

(3) SCL (Step 3): To evaluate the two-step learning approach in Step 3, we remove the unsupervised contrastive learning loss (UCL), resulting in more significant decreases of 10-17\% in average clustering scores across all three datasets. 

(4) Step 1 \& other strategies (K-Means++, UCL, MSE): Since the high-quality sampling strategy (Step 2) works in conjunction with multimodal representation learning (Step 3), we experiment with alternative strategies and observe their performance. Initially, applying K-Means++ directly after Step 1 leads to dramatic drops of over 10\% across all three datasets. Then, implementing UCL after Step 1 still brings noticeable decreases of 1.96\%, 5.93\%, and 6.07\% in average clustering metric scores. Lastly, we apply a mean squared error (MSE) loss between each sample feature and its corresponding cluster centroid, resulting in extremely low performance with decreases of 22.58\%, 11.17\%, and 15.91\% in average clustering metric scores. These ablation studies further validate the effectiveness of each component in our proposed UMC algorithm.

\subsection{Effect of the $K_{\text{near}}$ Selection Strategy}
In Section~\ref{sample_selection}, we introduce an automatic method for determining the optimal $K_{\text{near}}$ for each cluster. To demonstrate its efficacy, we compare it with a fixed $K_{\text{near}}$ approach, where the fixed value varies from 5 to 25 in increments of 5. We then evaluate the performance using ARI scores. 

As shown in Figure~\ref{ablation_k}, the automatic $K_{\text{near}}$ selection strategy outperforms all fixed $K_{\text{near}}$ settings (i.e., 1–2\% on MIntRec, 4–6\% on MELD-DA, and 6–9\% on IEMOCAP-DA). The reason is that the fixed strategy struggles with the imbalanced data distribution across clusters of varying sizes, whereas our approach adapts $K_{\text{near}}$ to the unique characteristics of each cluster. Importantly, this approach obviates the need for extensive manual hyper-parameter tuning while still ensuring excellent performance.

\subsection{Visualization}
\label{section: visualization}

Figure~\ref{tsne} uses t-SNE~\cite{van2008visualizing} to visualize representations on the IEMOCAP-DA dataset, with additional results provided in Appendix~\ref{appendix:representation_visualization}. SCCL exhibits substantial overlap among intent classes. CC displays more compact clusters, yet still presents implicit cluster boundaries. USNID shows clear cluster boundaries, but the different clusters are close in the feature space and difficult to discern. UMC (Text) demonstrates the most distinct cluster boundaries among text-based baselines, highlighting the robustness of the representations learned through our clustering algorithm. When incorporating non-verbal modalities, the multimodal representations learned by UMC reveal that each cluster is both compact and well-separated from others, verifying its efficacy.

\section{Conclusions}
This paper introduces the multimodal semantics discovery task and proposes a novel unsupervised multimodal clustering (UMC) method to address this critical challenge. UMC effectively utilizes non-verbal modalities for semantics discovery by constructing positive multimodal data augmentations. Besides, it proposes a novel high-quality sample selection mechanism and a two-step representation learning strategy. 

We conduct extensive experiments on both multimodal intent and dialogue act benchmark datasets. UMC achieves remarkable improvements of 2-8\% in standard clustering metrics compared to state-of-the-art clustering algorithms. Further analyses demonstrate the effectiveness of each component and the robustness of the learned representations conducive to clustering. We believe this work makes significant progress in this area and provide a solid foundation for related research. 

\section{Limitations}
There are two limitations in this work. Firstly, given the complexity of real-world multimodal intent datasets, the achieved clustering performance still suggests significant potential for further improvements. Secondly, while this study establishes a foundational approach for automatically determining the $K_{\text{near}}$  parameter, there is scope for exploring diverse methodologies within this automatic selection mechanism.

\section{Acknowledgements}
This work was supported by the National Natural Science Foundation of China (Grant No. 62173195), the National Science and Technology Major Project towards the new generation of broadband wireless mobile communication networks of Jiangxi Province (03 and 5G Major Project of Jiangxi Province) (Grant No. 20232ABC03402), High-level Scientific and Technological Innovation Talents "Double Hundred Plan" of Nanchang City in 2022 (Grant No. Hongke Zi (2022) 321-16), and Natural Science Foundation of Hebei Province, China
   (Grant No. F2022208006).

\bibliography{custom}
\bibliographystyle{acl_natbib}

\appendix

\section{Limitations of Text-only Information in Multimodal Semantics Discovery}
\label{appendix:example}
Multimodal information is crucial for semantics discovery, as it encompasses a broader range of communicative cues beyond mere text, such as tone of voice, facial expressions, and body language. These cues significantly enhance human communication by conveying subtle nuances and emotions that text alone cannot fully capture.

To illustrate this, we analyze examples from the MIntRec dataset, particularly focusing on two clusters with intents of \textit{Joke} (top) and \textit{Prevent} (bottom), as detailed in Table~\ref{appendix:table_example}. In the \textit{Joke} cluster, the first two examples contain text utterances with exaggerated rhetoric, clearly conveying the intent of \textit{Joke}. However, the latter four examples, with the semantics of \textit{Statement-opinion}, \textit{Question}, and \textit{Statement-non-opinion}, are less straightforward. Their real intention become clearer when considering non-verbal cues like exaggerated body language and expressions in a relaxed and happy tone.

Similarly, in the \textit{Prevent} intent cluster, the first two examples with clear negative directives are easily distinguished from text. However, the following three examples, which misleadingly suggest intentions of \textit{Agree}, \textit{Oppose}, and \textit{Inform} when relying solely on text. Here, non-verbal cues like \textit{nodding} and \textit{arm blocking} from body language, combined with a resolute voice of tone, are vital for discovering the real \textit{Prevent} intention.

Hence, incorporating non-verbal modalities is essential in real-world contexts for a comprehensive understanding of the complex semantics in human language. This shows the importance of leveraging non-verbal modalities when performing semantics discovery. 

\begin{table*}[!t] \tiny
    \centering
    \begin{tabular}{m{2.5cm}<{\centering} m{3cm}<{\centering} m{3cm}<{\centering} m{3cm}<{\centering} m{2cm}<{\centering}}
    \toprule
        Text
        & Video
        & Audio
        & Require Non-verbal Modalities & Useful Signals \\
        \midrule 
        he's, like, a major fox.
        & \includegraphics[scale=0.055]{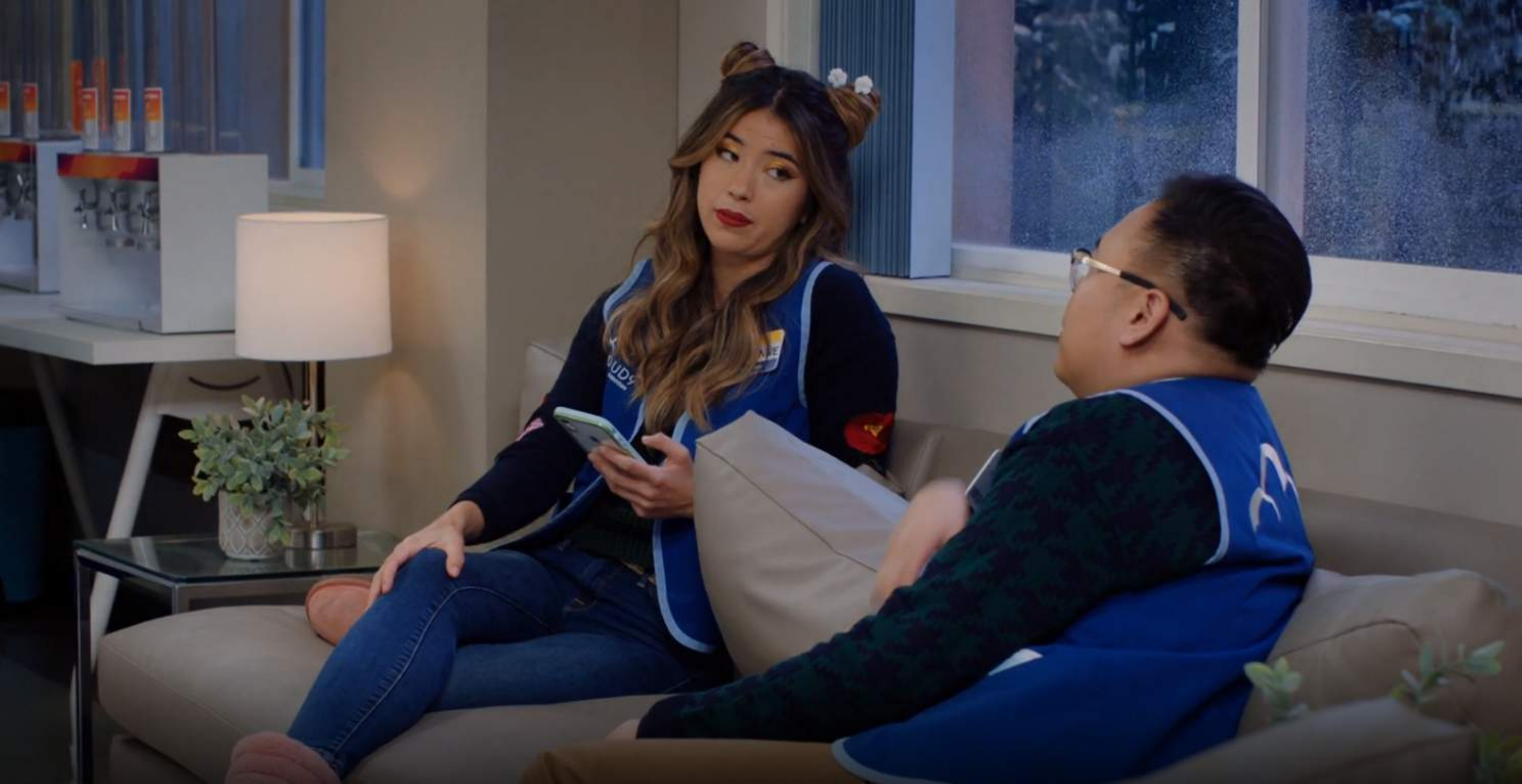} &\includegraphics[scale=0.3]{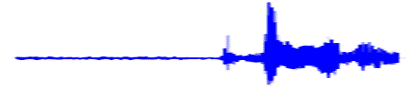} 
        & \scalebox{1.3}{\XSolidBold} & Natural Language \\

        you're like one of those monks in tibet.
        & \includegraphics[scale=0.055]{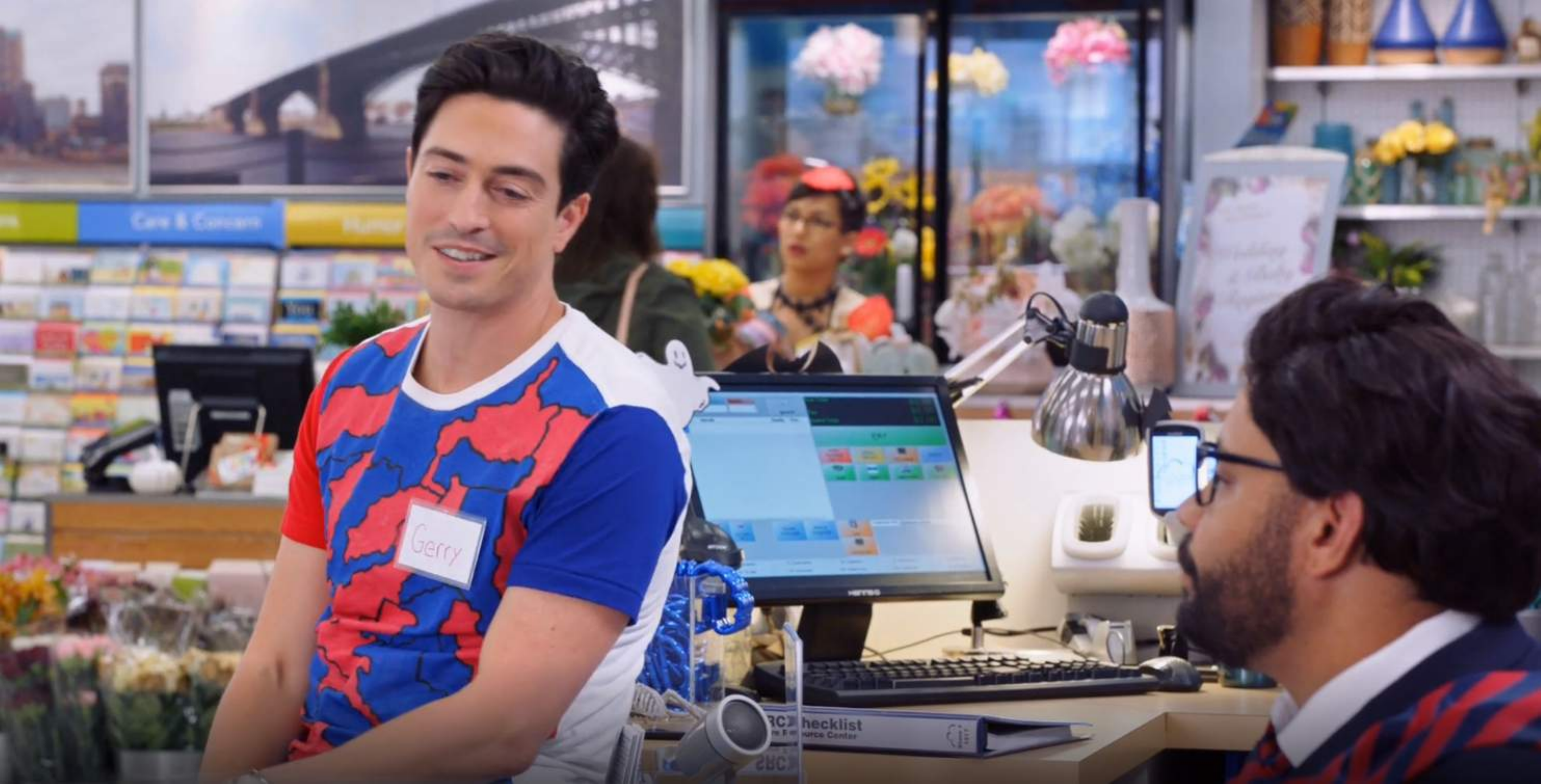} &\includegraphics[scale=0.3]{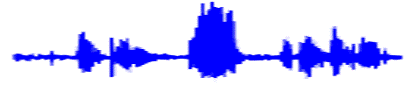} 
        & \scalebox{1.3}{\XSolidBold} & Natural Language \\

         and you're on the phone.
        & \includegraphics[scale=0.05]{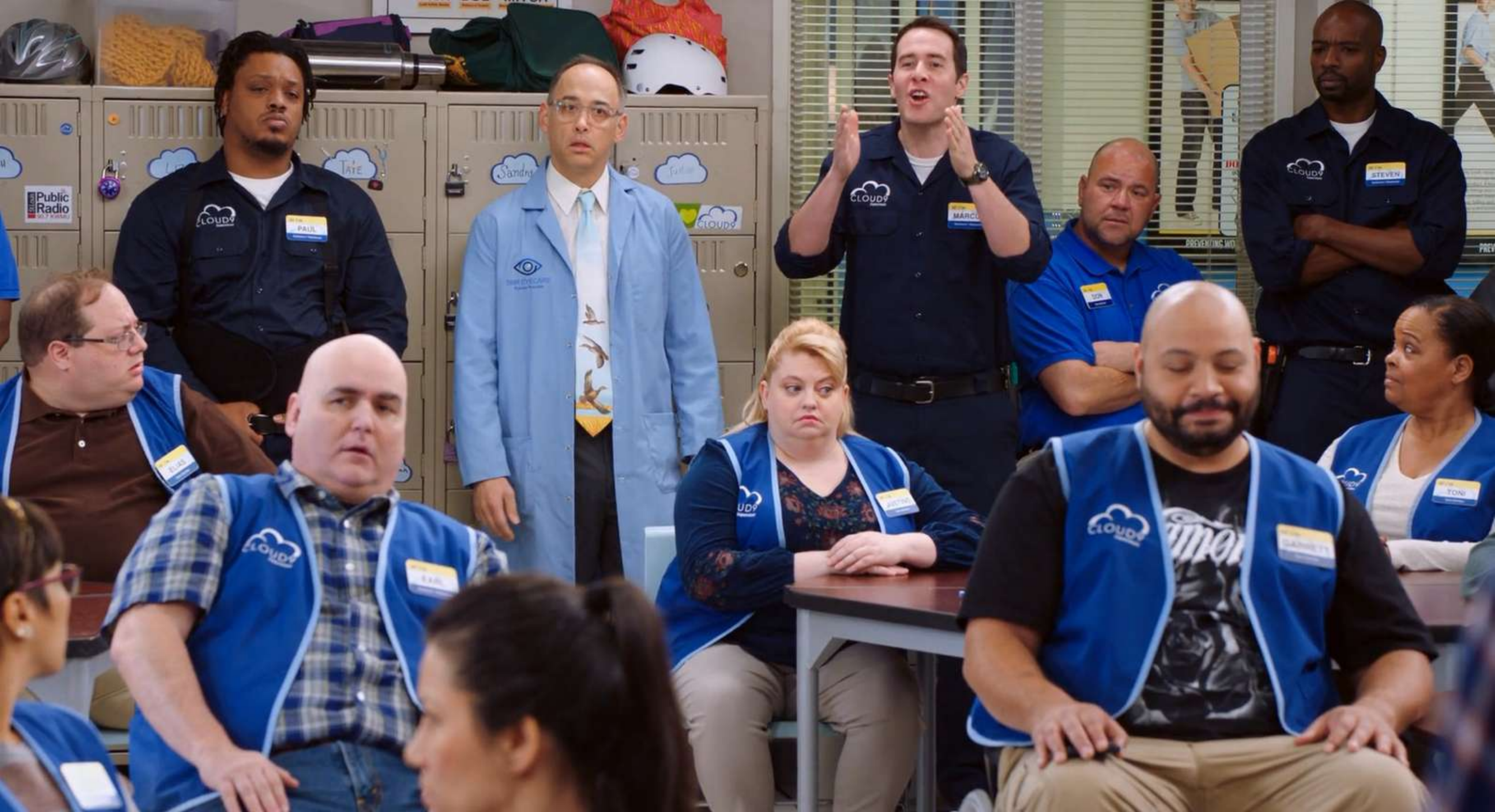} &\includegraphics[scale=0.3]{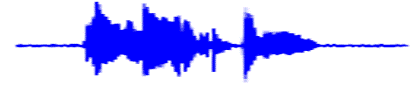} 
        & \scalebox{1.3}{\CheckmarkBold} & Tone of Voice, Expressions  \\

        and you got that from pants?
        & \includegraphics[scale=0.05]{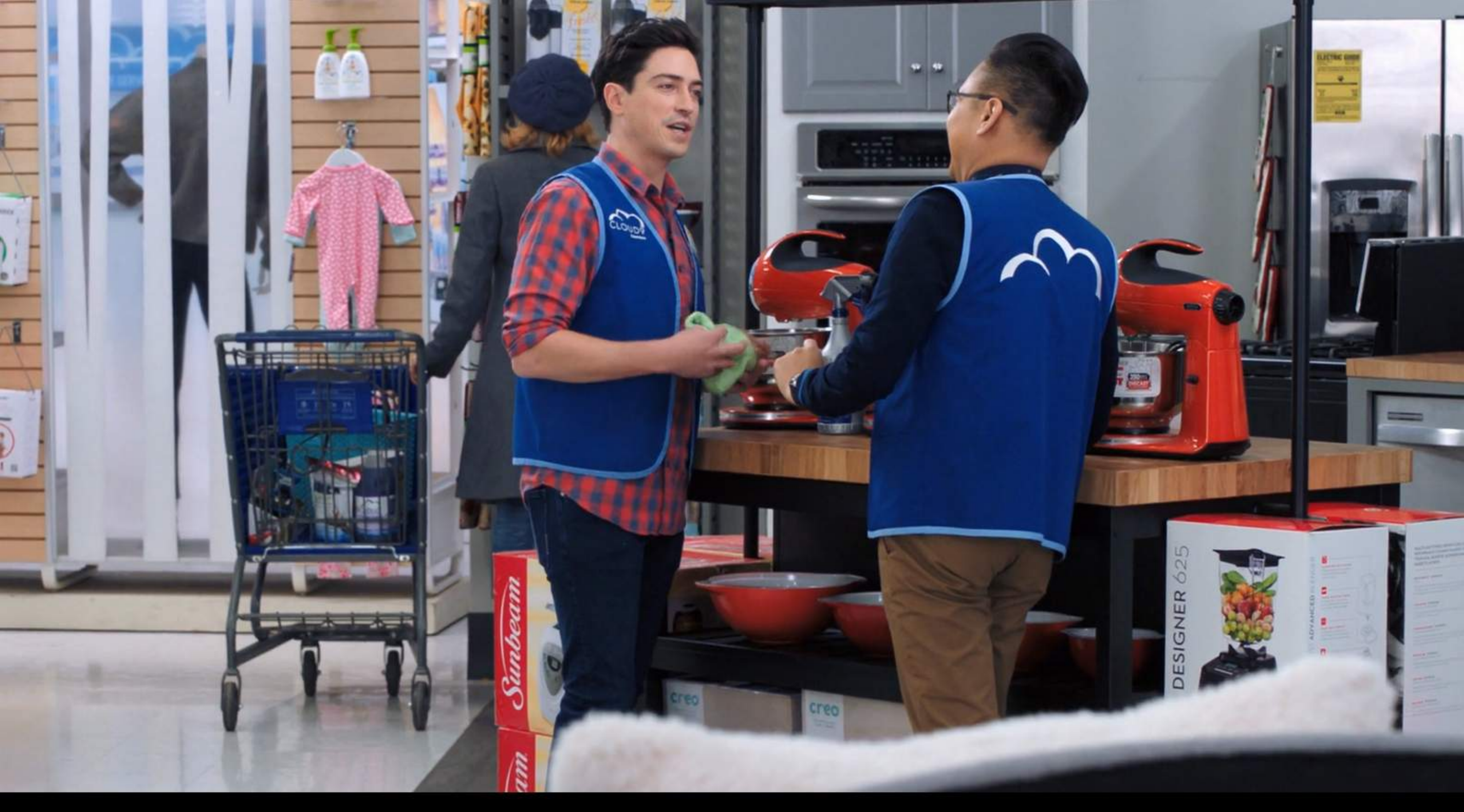} &\includegraphics[scale=0.3]{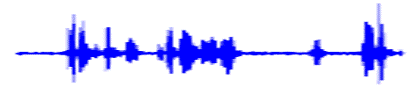} 
        & \scalebox{1.3}{\CheckmarkBold} & Tone of Voice, Expressions  \\

         running hard, water bad.
        & \includegraphics[scale=0.05]{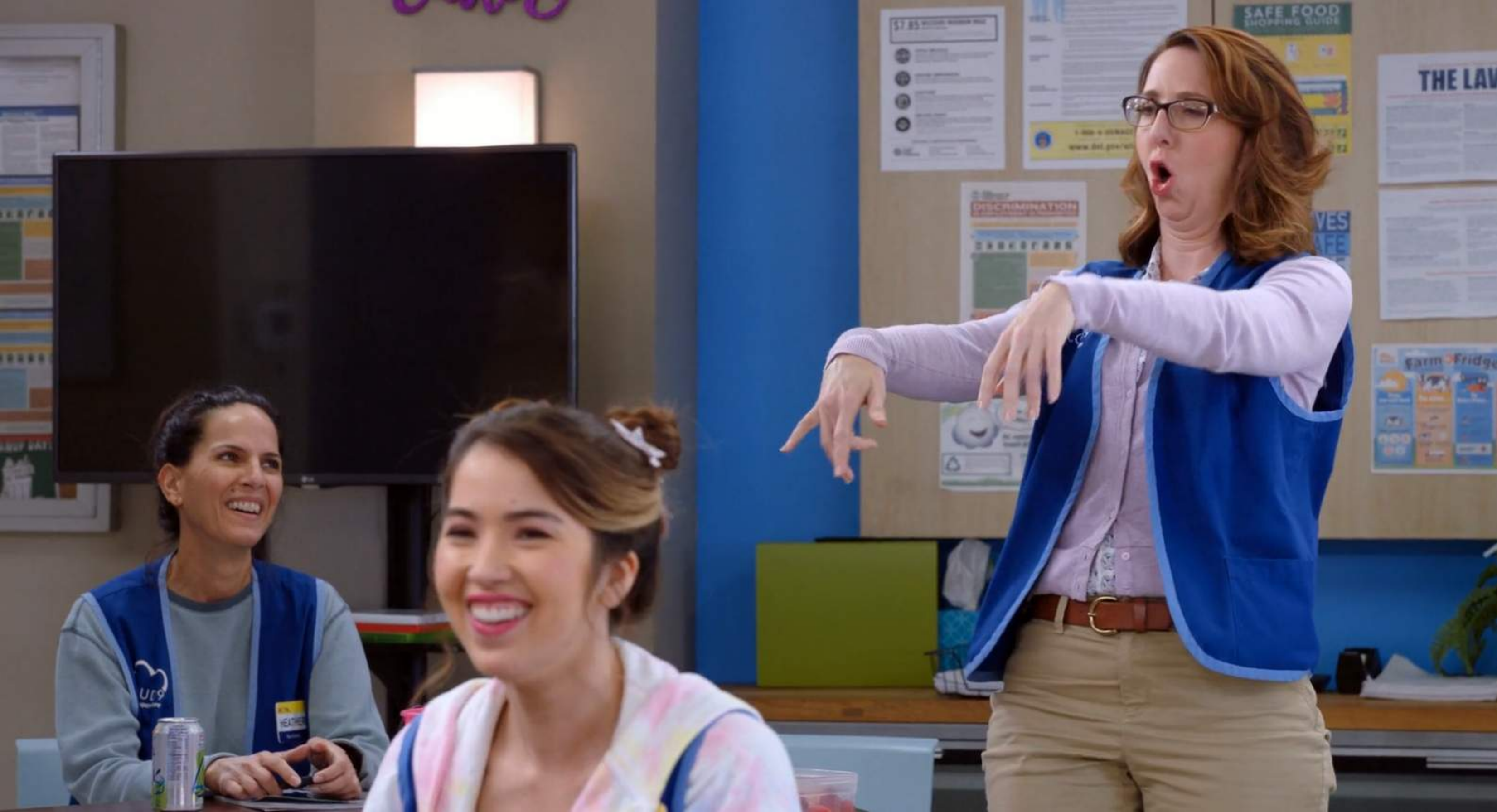} &\includegraphics[scale=0.3]{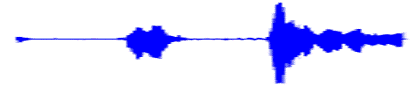} 
        & \scalebox{1.3}{\CheckmarkBold} & Body Language, Expressions \\

        i can do impressions too.
        & \includegraphics[scale=0.05]{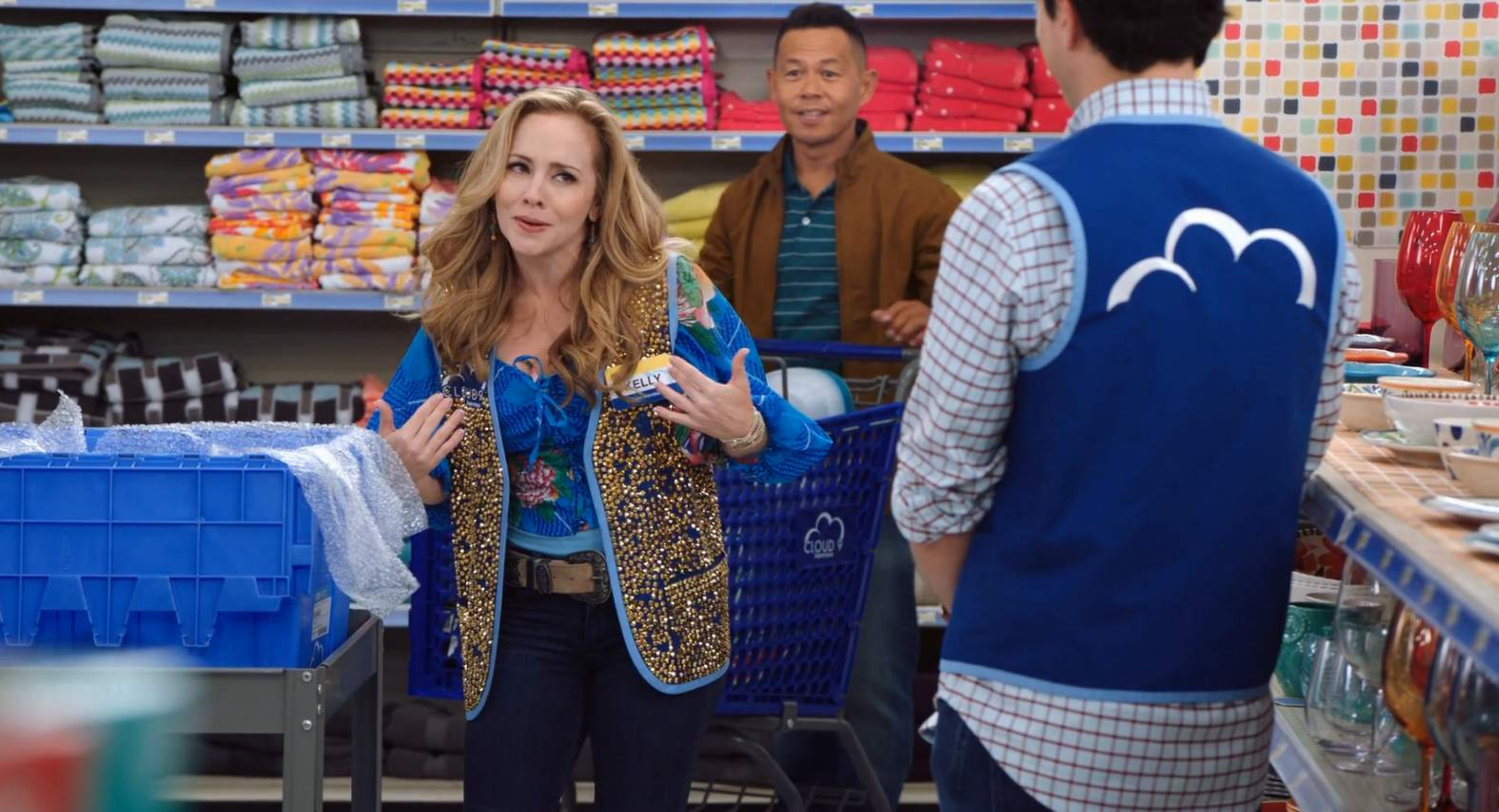} &\includegraphics[scale=0.3]{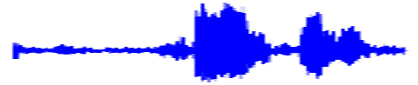} 
        & \scalebox{1.3}{\CheckmarkBold}  & Tone of Voice, Expressions\\
        
        \midrule
        \midrule

        okay, bo, stop. all right?
        & \includegraphics[scale=0.055]{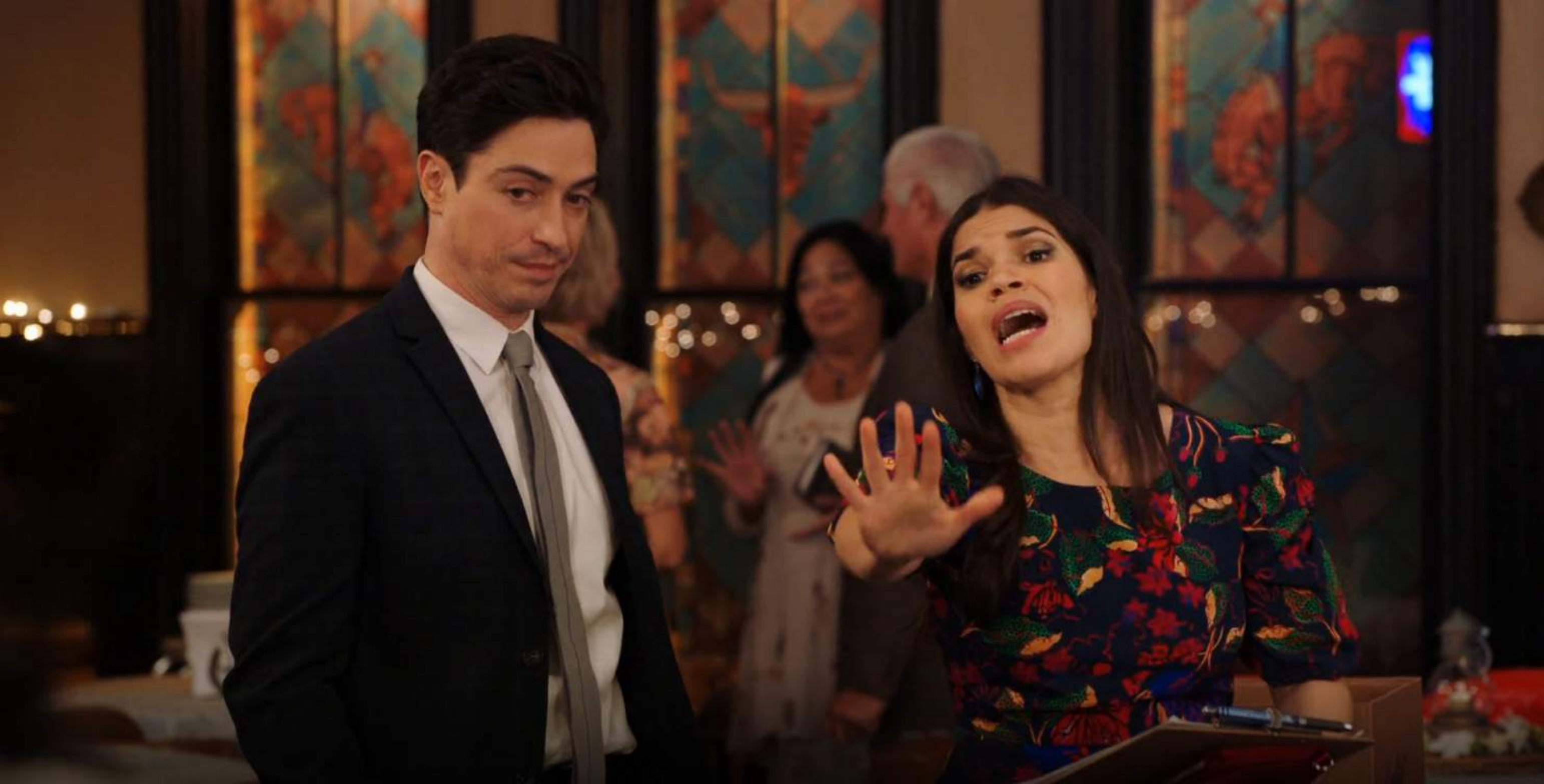} &\includegraphics[scale=0.3]{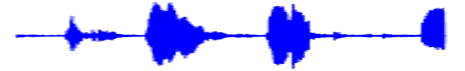} 
        & \scalebox{1.3}{\XSolidBold} & Natural Language  \\
        oh, god. sandra, stop, please.
        & \includegraphics[scale=0.055]{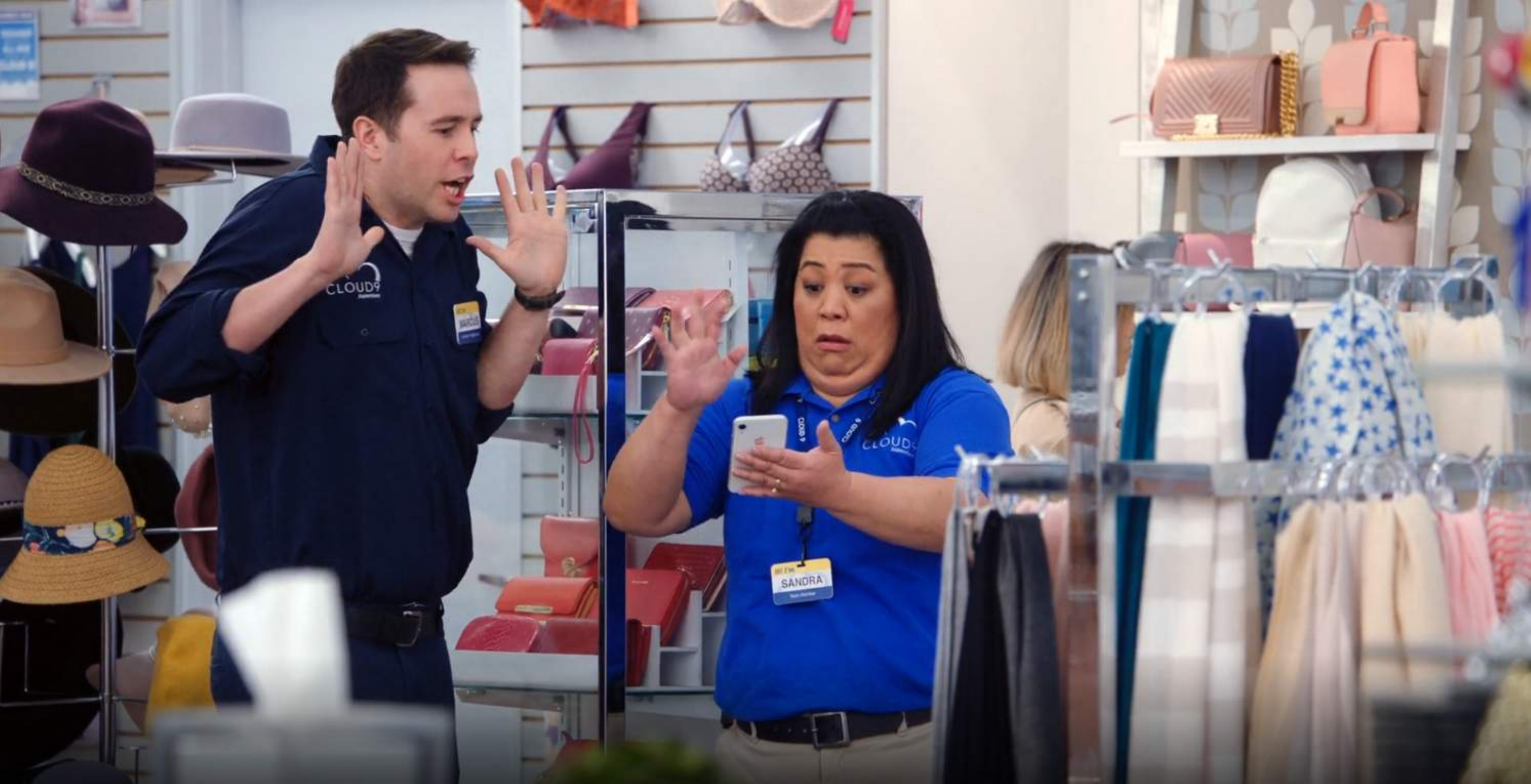} &\includegraphics[scale=0.3]{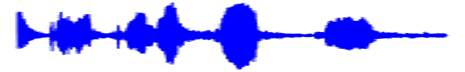} 
        & \scalebox{1.3}{\XSolidBold} & 
        Natural Language\\
        oh, yeah, yeah, yeah, yeah, yeah, yeah, yeah.
        & \includegraphics[scale=0.055]{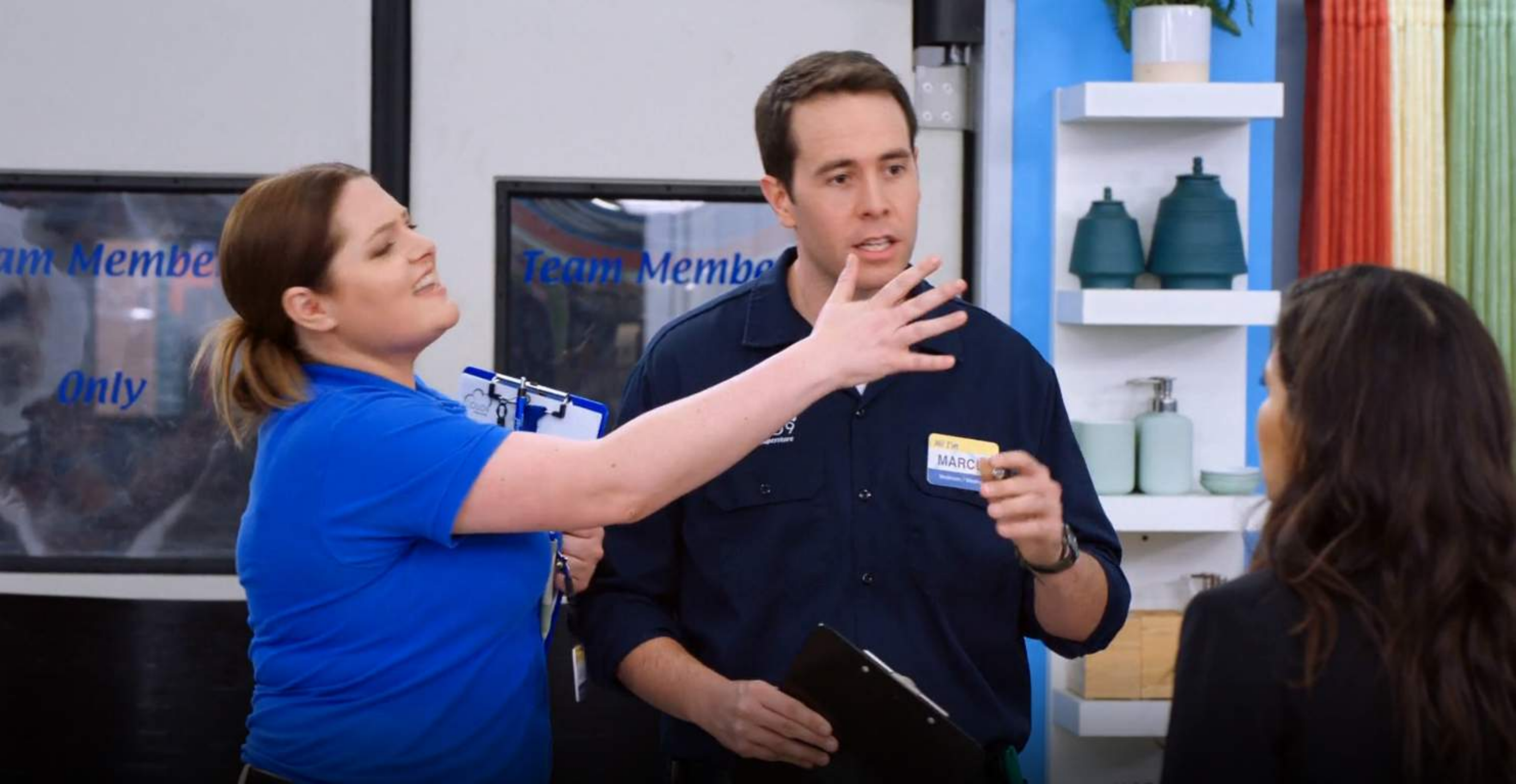} &\includegraphics[scale=0.3]{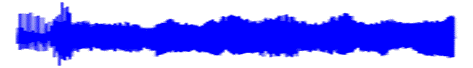} 
        & \scalebox{1.3}{\CheckmarkBold}  & Body Language\\
        oh, absolutely not.
        & \includegraphics[scale=0.055]{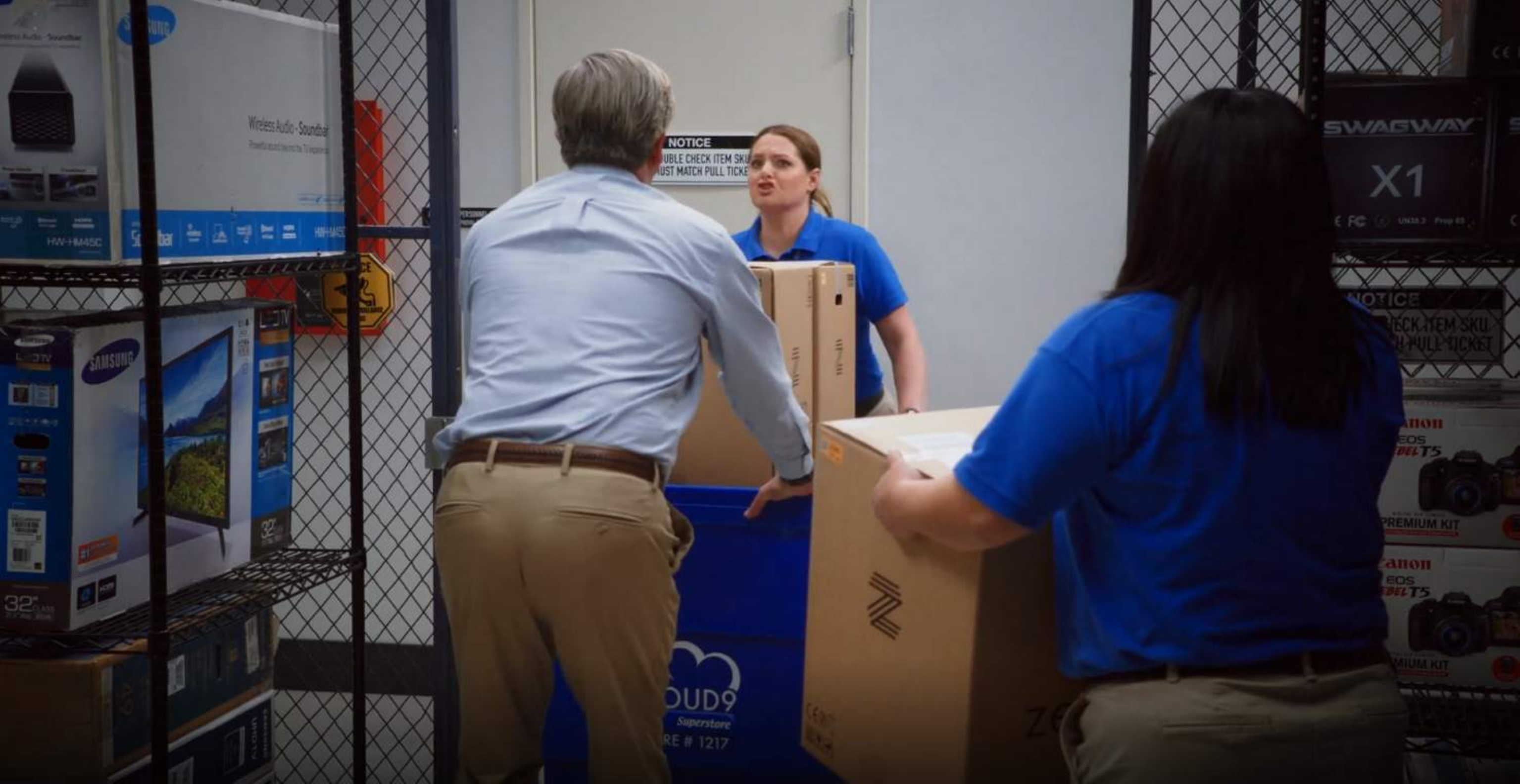} &\includegraphics[scale=0.3]{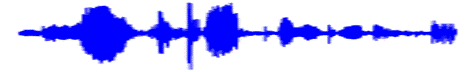} 
        & \scalebox{1.3}{\CheckmarkBold} & Body Language, Natural Language  \\
        mom, come on.
        & \includegraphics[scale=0.055]{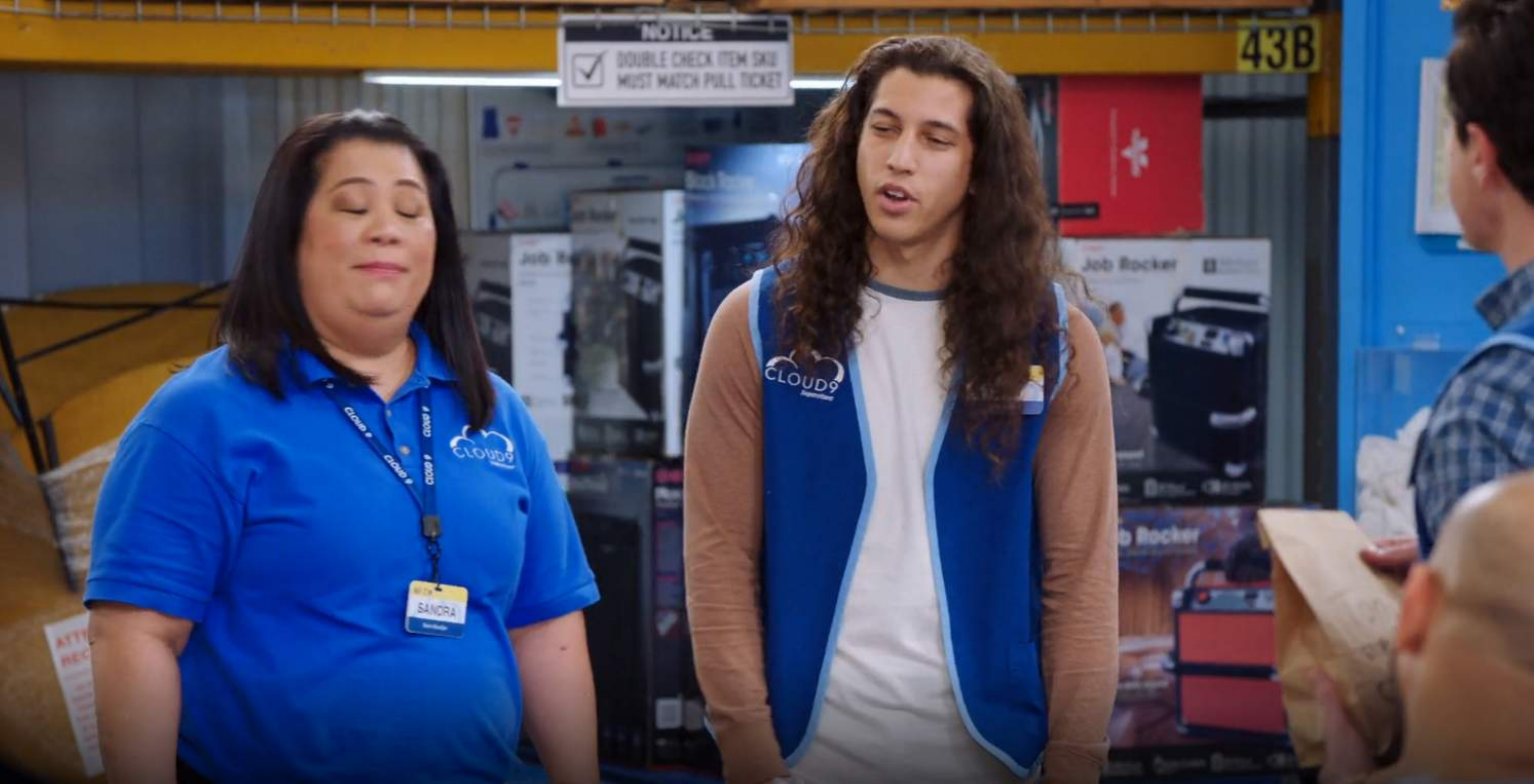} &\includegraphics[scale=0.3]{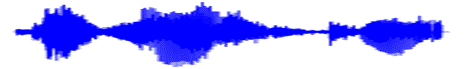} 
        & \scalebox{1.3}{\CheckmarkBold}  & Expressions, Tone of Voice\\
        \bottomrule
    \end{tabular}
     \caption{Real-world examples of $\textit{Joke}$ (top) and $\textit{Prevent}$ (bottom) intent clusters.}
     \label{appendix:table_example}
\end{table*}

\section{Applications of Multimodal Semantics Discovery}
\label{appendix:applications_future}

\textbf{Video Content Recommendation}: Online short video platforms, such as TikTok, have become globally popular, featuring content that includes text, video, and audio elements provided by content creators. Given the vast volume of videos on the internet, accurately tagging each video to match individual user preferences can be prohibitively expensive. Therefore, discovering potential user intentions from unsupervised multimodal data is crucial. An effective multimodal clustering method can discover user needs and group similar content, significantly improving the relevance of recommendations for content retrieval and search.

\textbf{Efficient Multimodal Data Annotation}: A well-trained multimodal clustering model is invaluable for processing real-world multimodal data. It can quickly create clusters based on similar multimodal characteristics, facilitating the identification and analysis of new patterns. Moreover, it enables the efficient generation of semantic annotations at the cluster level, speeding up the annotation process and reducing the workload compared to instance-level annotation.

\textbf{Virtual Human}: Virtual humans hold significant commercial value for many businesses, with some companies promoting custom-designed robots as flagship products. However, effective virtual humans must be able to accurately capture human intentions from various signals, including natural language, body language, facial expressions, and vocal tone. Given that data from real-world human-machine interactions are often unsupervised, it is vital for virtual humans to discern potential user needs from clustered data. This capability allows them to offer better performance and interact with humans in a more natural and fluent manner.

Overall, multimodal semantics discovery opens up new possibilities for the analysis and interpretation of unsupervised multimodal data, which is increasingly prevalent in our digital communication era.

\section{Additional Related Works}
\label{appendix:additional_related_works}
\subsection{Multi-view Clustering}

Multi-view clustering primarily employs matrix optimization algorithms such as CDD~\cite{huang2021cdd}, COMIC~\cite{peng2019comic}, OS-LF-IMVC~\cite{zhang2021one}, and SMVSC~\cite{sun2021scalable}. These algorithms utilize graphical or spatial methods to mathematically divide clustering into several sub-tasks and then iteratively optimize the subtask matrices. However, multi-view clustering may become inefficient when processing high-dimensional data, and its time cost can increase at an ultra-linear rate with larger datasets. Besides, the design of optimization objectives in multi-view clustering methods presents certain challenges and does not always guarantee favorable results.

In contrast, multimodal clustering, which tends to focus on deep neural network methodologies, can alleviate these difficulties. For example, XDC~\cite{alwassel2020self} clusters two separate modalities and employs cross-modal pseudo-labels as a supervisory signal for model training, effectively utilizing the semantic connections and distinctions between different modalities. DMC~\cite{hu2019deep} uses an exponential function approximation to enable differentiable minimum optimization for clustering, drawing data points closer to their cluster center. It is important to note that these methods are limited to bimodal learning rather than accommodating multiple modalities.

\begin{table}[!t] \scriptsize
    \centering
    \begin{tabular}{@{\extracolsep{0.01pt}}c|l|cccccc}
    \toprule
         & Backbones & NMI & ARI & ACC & FMI   \\
        \midrule
        & ResNet-50 + wav2vec 2.0 & 47.90 & 23.04 & 42.61 & 27.79  \\
        & Swin Transformer + WavLM & \textbf{49.46} & \textbf{24.79} & \textbf{44.00} & \textbf{29.48} \\
        \bottomrule
    \end{tabular}
    \caption{\label{appendix:table_effect_of_multimodal} Effect of the multimodal features on the MIntRec dataset.}
    \vspace{-0.5cm}
\end{table}
\begin{figure*}[!t]
    \centering
    \includegraphics[scale=0.2]{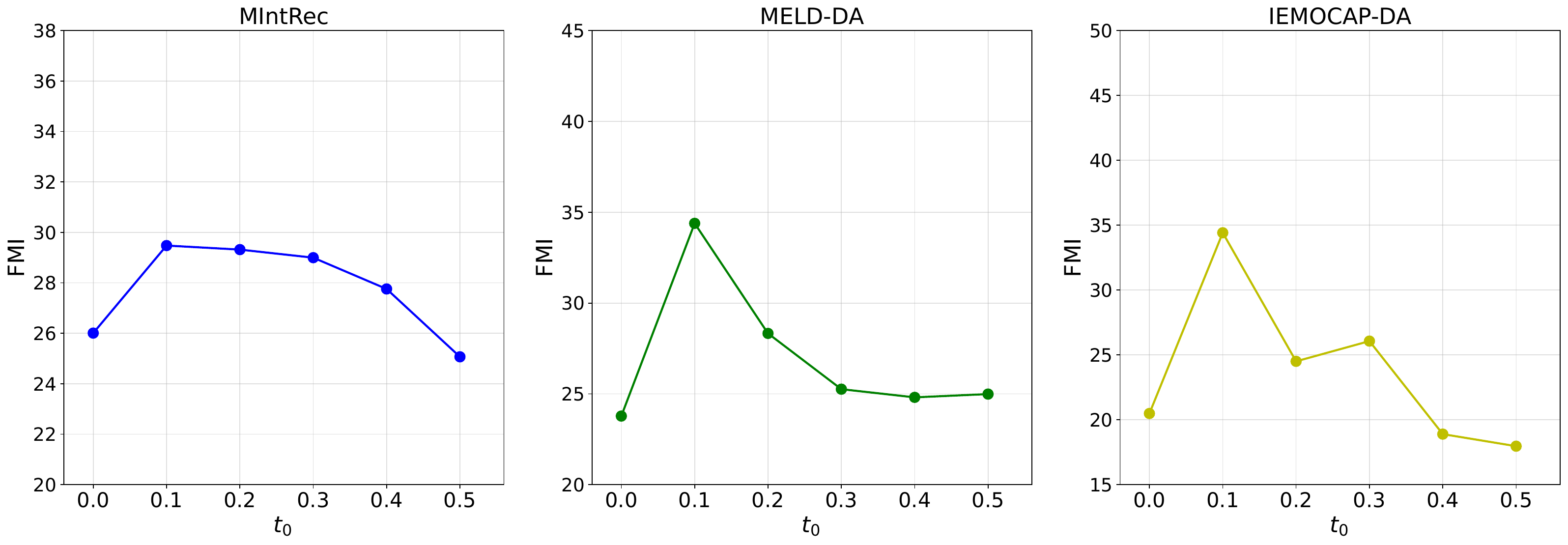}
    \caption{Results of clustering with varying initial thresholds $t_{0}$.}
    \label{ablation_t0}
\end{figure*}
\begin{figure*}[!t]
    \centering
    \includegraphics[scale=0.2]{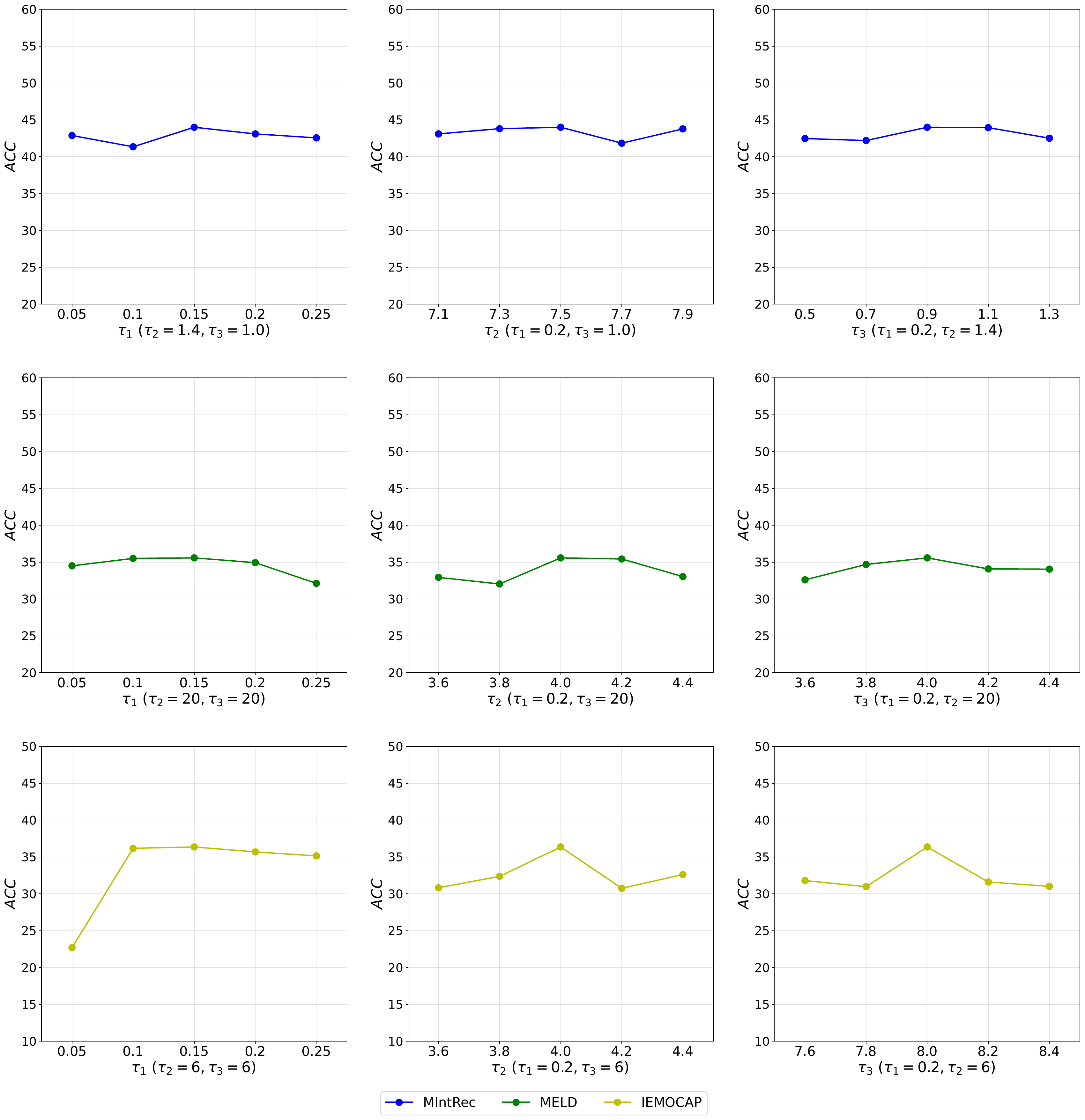}
    \caption{Sensitivity analysis of $\tau_{1}$, $\tau_{2}$, and $\tau_{3}$ on the three datasets.}
    \label{appendix:figure_sensitivity_analysis}
    % \vspace{-0.2cm}
\end{figure*}
\subsection{Multimodal Language Analysis}
Multimodal language analysis has introduced numerous datasets~\cite{zadeh2016mosi,zadeh2018multimodal,yu2020ch} and multimodal fusion methods~\cite{tsai2019multimodal,sun2020learning,rahman2020integrating,hazarika2020misa,yu2021learning,han2021improving,maharana2022multimodal,unimse,wei-tackling,yang-confede,multiemo}. 

While most research has focused on meta properties like emotions or sentiment, less attention has been given to the content semantics of multimodal utterances.~\citet{zhou2024token} has specifically designed a method for multimodal intent recognition, leveraging the text modality to guide the learning of prompts from non-verbal modalities. However, this method is inapplicable for unsupervised scenarios.  

To address this, EMOTyDA~\cite{EMOTyDA} offers dialogue act labels that complement two multimodal emotion datasets~\cite{busso2008iemocap,MELD}, and recent studies have ventured into multimodal dialogue act classification~\cite{saha-etal-2021-towards_,saha2021emotion}.~\citet{maharana2022multimodal} introduce a  dataset for recognizing operational intents in instructive videos, employing a multimodal cascaded cross-attention late fusion model. MIntRec~\cite{zhang2022mintrec} provides the first  multimodal dataset for conversational intent recognition, using multimodal fusion methods as benchmarks. However, these works depend on supervised learning with provided labels, with few studies in the area of unsupervised multimodal language analysis. Very recently,~\citet{zhang2024mintrec} introduces the first large-scale multimodal dataset for both intent recognition and out-of-scope detection in conversations, highlighting the challenges of existing machine learning methods in understanding complex semantics within multimodal utterances.

\begin{figure}[!t]
    \centering
    \includegraphics[scale=0.18]{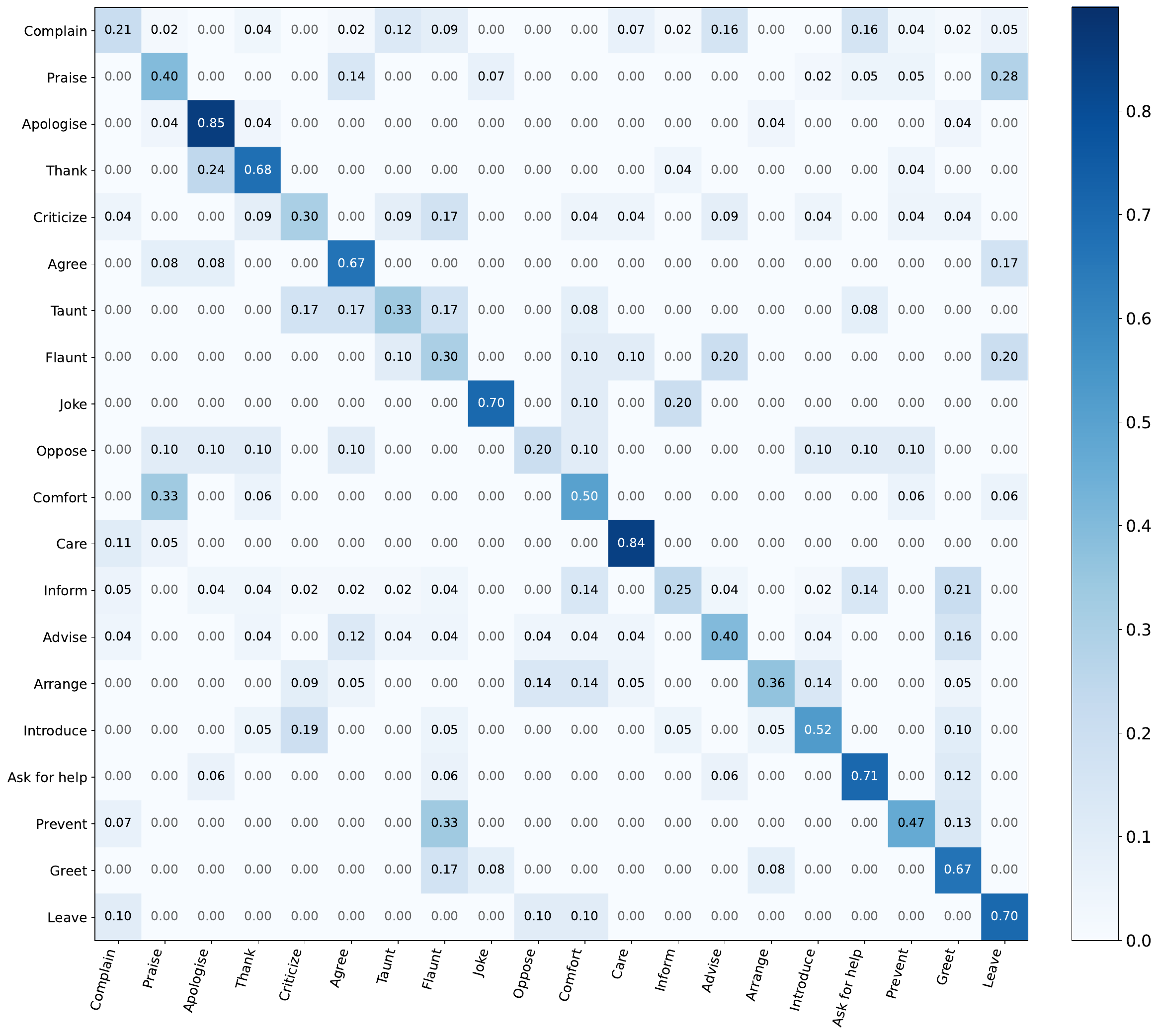}
    \caption{Confusion matrix on the MIntRec dataset.}
    \label{appendix:figure_error_analysis}
\end{figure}
\section{Effect of Multimodal Features}
\label{appendix:effect_of_multimodal}
In this study, we select the Swin Transformer and WavLM, two state-of-the-art models in computer vision and speech signal processing, as backbones for extracting multimodal features. These models demonstrate superior performance compared to the original features used in the MIntRec~\cite{zhang2022mintrec} and EMOTyDA~\cite{EMOTyDA} datasets. Due to the unavailability of features from EMOTyDA, which poses a challenge for reproduction, we employed features from the MIntRec dataset. MIntRec utilizes ResNet~\cite{He_2016_CVPR} and wav2vec 2.0~\cite{baevski2020wav2vec} for video and audio modalities, respectively.

As illustrated in Table~\ref{appendix:table_effect_of_multimodal}, our results show that Swin Transformer and WavLM enhance performance by over 1\% on the MIntRec dataset. This improvement evidences their effectiveness in modeling multimodal representations and capturing the semantics of non-verbal modalities, which are pivotal for cross-modal interactions.

\section{Selection of $t_0$}
\label{appendix:t_0_selection}
To select the appropriate parameter $t_0$ as mentioned in Eq.~\ref{eqt}, we vary $t_0$ from 0.0 to 0.5 at intervals of 0.1 and present the experimental results of the clustering metric FMI on three multimodal intent datasets.

As shown in Figure~\ref{ablation_t0}, the clustering results fluctuate with different values of $t_0$, with $t_0$=0.1 generally achieving the best performance. Specifically, this value of $t_0$ achieve the highest performance on the MIntRec and IEMOCAP-DA datasets, and comparable performance on the MELD-DA dataset. This is reasonable because a larger $t_0$ tends to include more data initially, which may introduce more low-quality data as anchors, thereby hindering the learning of representations conducive to effective clustering.

\begin{figure*}[!th]
    \centering
    \includegraphics[scale=0.15]{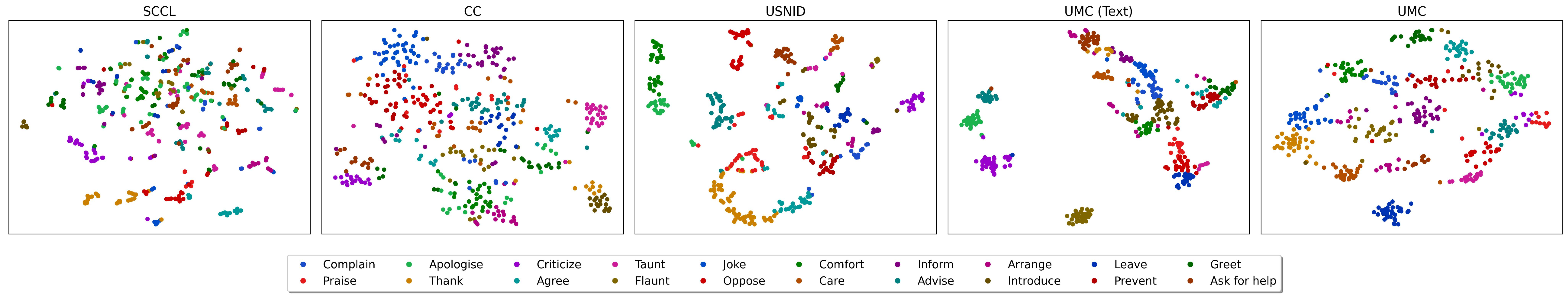}
    \caption{Visualization of representations on the MIntRec dataset.}
    \label{appendix:mintrec_visualization}
\end{figure*}

\begin{figure*}[!th]
    \centering
    \includegraphics[scale=0.15]{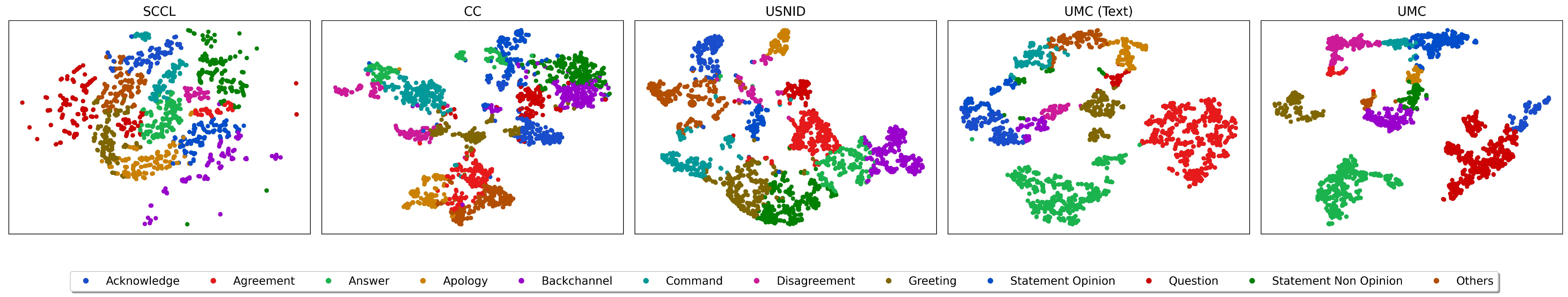}
    \caption{Visualization of representations on the MELD-DA dataset.}
    \label{appendix:meld_data_visualization}
\end{figure*}
\section{Dataset Specifications and Split Details}
\label{appendix:dataset_specifications}
\textbf{Multimodal Intent Dataset}: MIntRec~\cite{zhang2022mintrec} is the premier dataset for multimodal intent recognition in conversation scenarios, spanning text, audio, and video modalities. It comprises 20 intent classes with 2,224 high-quality annotated samples. The original dataset has a 3:1:1 split for training, validation, and testing. As unsupervised clustering does not require the validation set, we merge it with the training set, resulting in a 4:1 ratio between the training and testing sets.

\textbf{Multimodal Dialogue Act Datasets}: We use two large-scale multimodal dialogue act datasets, MELD-DA and IEMOCAP-DA, which are derived from the MELD~\cite{MELD} and IEMOCAP~\cite{busso2008iemocap} datasets, respectively. The EMOTyDA~\cite{EMOTyDA} dataset provides dialogue act labels for these datasets, encompassing 12 dialogue act classes. We maintain a 4:1 data split ratio for training and testing, consistent with the split used for MIntRec.

\section{Evaluation Metrics}
\label{em}
Particularly, ACC is calculated by aligning predictions and ground truth using the Hungarian algorithm, as described in~\cite{zhang2021supporting, USNID}. For NMI, ACC, FMI, the range of possible values is from 0 to 1, while ARI ranges from -1 to 1. Higher values of all these metrics indicate better clustering performance.

 The normalized mutual information (NMI) is defined as:
\begin{align}
\textrm{NMI}(\mathbf{y}^{gt}, \mathbf{y}^{p}) &= \frac{\it{MI}(\mathbf{y}^{gt}, \mathbf{y}^{p})}{\frac{1}{2} (H(\mathbf{y}^{gt}) + H(\mathbf{y}^{p}))},
\end{align}
where $\mathbf{y}^{gt}$ and $\mathbf{y}^{p}$ are the ground-truth and predicted labels, respectively. $\it{MI}(\mathbf{y}^{gt}, \mathbf{y}^{p})$ represents the mutual information between $\mathbf{y}^{gt}$ and $\mathbf{y}^{p}$, and $H(\cdot)$ is the entropy. The mutual information is normalized by the arithmetic mean of $H(\mathbf{y}^{gt})$ and $H(\mathbf{y}^{p})$, and the resulting NMI values fall within the range of [0, 1].

The adjusted Rand index (ARI) is defined as:
\begin{align}
&\textrm{ARI} = \nonumber \\ 
&\frac{
\sum_{i, j}\binom{n_{i, j}}{2}-[\sum_{i}\binom{u_{i}}{2}\sum_{j}\binom{v_{j}}{2}] / \binom{n}{2}
}
{
\frac{1}{2}[\sum_{i}\binom{u_{i}}{2}+\sum_{j}\binom{v_{j}}{2}]-[\sum_{i}\binom{u_{i}}{2}\sum_{j}\binom{v_{j}}{2}]/\binom{n}{2}
},
\end{align}
where $u_{i}=\sum_{j}n_{i,j}$, and $v_{j}=\sum_{i}n_{i,j}$. $n$ is the number of samples, and $n_{i,j}$ is the number of samples that have both the $i^{\textrm{th}}$ predicted label and the $j^{\textrm{th}}$ ground-truth label. The values of ARI fall within the range of [-1, 1].

The accuracy (ACC) is defined as:
\begin{align}
     \textrm{ACC}(\mathbf{y}^{gt}, \mathbf{y}^{p}) &=\max _m \frac{\sum_{i=1}^n \mathbb{I} \left\{y^{gt}_i=m\left(y^{p}_i\right)\right\}}{n},
\end{align}
where $m$ is a one-to-one mapping between the ground-truth label $\mathbf{y}^{gt}$ and predicted label $\mathbf{y}^{p}$ of the $i^{\textrm{th}}$ sample. The Hungarian algorithm efficiently obtains the best mapping $m$. The values of ACC range from [0, 1].

FMI (Fowlkes-Mallows Index) is defined as:
\begin{align}
\textrm{FMI}(\mathbf{y}^{gt}, \mathbf{y}^{p}) &= \frac{\textrm{TP}}{\sqrt{(\textrm{TP} + \textrm{FP})(\textrm{TP} + \textrm{FN})}},
\end{align}
where $\mathbf{y}^{gt}$ and $\mathbf{y}^{p}$ are the ground-truth and predicted labels, respectively. $\textrm{TP}$ represents the number of true positive instances, $\textrm{FP}$ is the number of false positive instances, and $\textrm{FN}$ is the number of false negative instances. The FMI is calculated as the ratio of true positive instances to the geometric mean of the product of false positives and false negatives. The values of FMI range from 0 to 1, with higher values indicating better clustering performance.

\section{Hyper-parameter Sensitivity Analysis}
\label{appendix:sensitivity_analysis}
We conduct a sensitivity analysis on the key hyper-parameters $\tau_1$, $\tau_2$, and $\tau_3$, which are essential for multimodal unsupervised pre-training and representation learning. The results are displayed in Figure~\ref{appendix:figure_sensitivity_analysis}.

Initially, with $\tau_2$ and $\tau_3$ set to their optimal values, we explore the impact of varying $\tau_1$ on multimodal clustering. The optimal value for $\tau_1$ is 0.15 across all three datasets, and any deviation from this value results in a performance decline. Next, keeping $\tau_1$ and $\tau_3$ constant, we find that the optimal values for $\tau_2$ are 7.5 for MIntRec and 4.0 for the MELD-DA and IEMOCAP-DA datasets, respectively, with slight performance fluctuations on the IEMOCAP-DA dataset. Finally, by fixing $\tau_1$ and $\tau_2$, we observe that the optimal values of $\tau_3$ are 0.9, 4.0, and 8.0 for the MIntRec, MELD-DA, and IEMOCAP-DA datasets, respectively. The performance also exhibits fluctuations on the IEMOCAP-DA dataset.

\section{Error Analysis}
\label{appendix:error_analysis}
Utilizing Hungarian alignment between predictions and ground truth, we present the confusion matrix in Figure~\ref{appendix:figure_error_analysis}. First, we note that several classes with simpler semantics, such as \textit{Apologise} and \textit{Care}, achieve an accuracy of 85\%. This is reasonable, as they can be easily identified using textual information alone, and the incorporation of non-verbal information maintains this high accuracy.

However, for moderately difficult intent classes like \textit{Agree}, \textit{Joke}, and \textit{Ask for help}, which achieve over 60\% accuracy, non-verbal cues such as \textit{nodding} and \textit{gestures} play a crucial role in inferring the true intent. On the other hand, some intent classes, including \textit{Oppose}, \textit{Complain}, \textit{Taunt}, and \textit{Flaunt}, are particularly challenging, with very few or even no samples being accurately clustered. These classes often exhibit complex semantics that require a sophisticated integration of different modalities to accurately interpret real human intentions, resulting in poor performance in multimodal clustering. The overall modest performance also indicates significant room for improvement in the field of unsupervised multimodal clustering.
\section{Representation Visualization}
\label{appendix:representation_visualization}
Besides the visualized representations on the IEMOCAP-DA dataset, as introduced in Figure~\ref{tsne}, we visualize the representations on the MIntRec and MELD-DA datasets. These are respectively illustrated in Figure~\ref{appendix:mintrec_visualization} and Figure~\ref{appendix:meld_data_visualization}. SCCL struggles to effectively learn cluster-level features, resulting in a trivial case of discrete points. CC performs better, forming cluster shapes that are relatively easy to distinguish. USNID shows more compact cluster boundaries, but the boundaries between adjacent clusters are still somewhat indistinct. UMC (Text) performs the best among the text baselines but still has difficult clusters that are close to others. In contrast, our proposed UMC method displays explicit decision boundaries between different clusters, with intra-cluster cohesion being more compact compared to the other methods.
    
\end{document}